# Can Artificial Intelligence Trade the Stock Market?

Jędrzej Maskiewicz[a], Paweł Sakowski[b*],

[a] *Quantitative Finance Research Group, Department of Quantitative Finance, Faculty of Economic Sciences, University of Warsaw, ul. Długa 44/50, 00-241 Warsaw, Poland.*

[b] *Quantitative Finance Research Group, Department of Quantitative Finance, Faculty of Economic Sciences, University of Warsaw, ul. Długa 44/50, 00-241 Warsaw, Poland. ORCID: https://orcid.org/0000-0003-3384-3795*

\* Corresponding author e-mail: p.sakowski@uw.edu.pl

**Abstract:** The paper explores the use of Deep Reinforcement Learning (DRL) in stock market trading, focusing on two algorithms: Double Deep Q-Network (DDQN) and Proximal Policy Optimization (PPO) and compares them with Buy and Hold benchmark. It evaluates these algorithms across three currency pairs, the S&P 500 index and Bitcoin, on the daily data in the period of 2019-2023. The results demonstrate DRL's effectiveness in trading and its ability to manage risk by strategically avoiding trades in unfavorable conditions, providing a substantial edge over classical approaches, based on supervised learning in terms of risk-adjusted returns.

# INTRODUCTION

In the rapidly evolving field of financial technology, algorithmic trading has taken center stage due to its ability to execute orders at high speed and with precise control. Utilising complex algorithms to analyse market conditions, algorithmic trading now accounts for around 70% of all equities traded in the US alone and as high as 90% for the Foreign Exchange Market (FOREX) (Kissel, 2020). As we look to the future, Deep Learning (DL) is on the way to revolutionise the landscape of algorithmic trading. This branch of machine learning, characterised by its ability to process vast amounts of data through Neural Networks (NN) with multiple layers of abstraction, excels in identifying subtle patterns and correlations that traditional algorithms might miss. Further advancing this frontier is Deep Reinforcement Learning (DRL), arguably today's most sophisticated family of algorithms. By combining the analytical strength of DL with the adaptive decision-making processes of the Reinforcement Learning (RL), DRL enables optimisation actions based on cumulative experience, excelling in dynamic and uncertain environments[1].

The focus of this paper is therefore that DRL, as an advanced form of artificial intelligence, can effectively conduct financial market trading by autonomously identifying and exploiting patterns and relationships within complex, high-dimensional data. The primary goal of this research is to test this hypothesis by evaluating the impact of different NN architectures on the performance of trading algorithms in real-world trading scenarios. Specifically, we focus on two advanced DRLs: Double Deep Q-Network (DDQN) (Mnih et al., 2015) and Proximal Policy Optimisation (PPO) (Schulman et al., 2017) and two network architectures—the Fully Connected Neural Network (Williamset et al., 1989) and the Transformer network (Vaswani et al., 2017).

This task is particularly challenging due to the inherent complexities and dynamic nature of financial markets, where multiple factors such as economic indicators, geopolitical events, and exchange sentiment can drastically affect outcomes. Moreover, the volatile nature of these markets demands that trading algorithms learn how to execute trades effectively and adapt swiftly to unforeseen changes, rather than merely predict market movements. The focus is on enabling the agent to autonomously learn and refine trading strategies that respond appropriately to market conditions, thereby optimising performance and enhancing decision-making processes. DRL should excel at adapting to market realities, various trends, and different volatilities.

---

[1] To explore further the innovative use of deep reinforcement learning in algorithmic trading, see the detailed review by Pricope (2021).

One of the significant challenges in developing effective trading strategies lies in the ability to discover and validate investment strategies that are robust across various assets, time frames, and market conditions. Traditional approaches often fail to generalise well enough only to the conditions they were tested under, leading to suboptimal strategies when market dynamics changes. To address these issues, this paper will leverage the concept of moving forward optimisation, which continuously evaluates DRL strategy using recent data, improving strategy responsiveness and effectiveness over future. Furthermore, to test the reliability of DRL in algorithmic trading, we would assess strategy over different asset classes with their own characteristics. These will include major currency pairs, which represent the largest and most liquid segment of the FOREX market; the S&P 500 index, a benchmark for U.S. equities; and Bitcoin, a leading cryptocurrency known for its high volatility. This should test the versatility and adaptability of DRL strategies.

Therefore, the structure of the paper is as follows: the upcoming section will lay the groundwork for understanding Reinforcement Learning with its core fundamentals. Additionally this chapter presents relevant literature and research on Efficient Market Hypothesis, Reinforcement Learning and Deep Reinforcement Learning. After this, we will delve into the essential statistical and mathematical foundations of Reinforcement Learning, Deep Learning and Deep Reinforcement Learning, focusing on their key components. Next, we will transition our discussion to the practical methodology of our research, including evaluation approaches. This functional section would be followed by a presentation of the results derived from our analysis. A subsequent chapter will evaluate these findings and offer insights. Finally, we will provide comprehensive recommendations for future work, specifically tailored to the application of DRL agents in stock market trading.

# CHAPTER I

# Background and related work

## 1.1. Key Concepts in Reinforcement Learning

Reinforcement Learning (RL) is a branch of machine learning where an algorithm, referred to as an **agent**, learns to make decisions by interacting with a dynamic **environment**. Through these interactions, the agent performs **actions** $a$ and receives **rewards** $r$ (or penalties as negative rewards) based on its actions. The decisions are chosen by a **policy** $\pi$, which is a strategy that maps environment **states** $s$ to actions that maximise long-term rewards. In the case of algorithmic trading: action $a$ would be a position taken, environment a stock market, state $s$ would be an input (OHLC data, trading indicators, current position, etc.) and reward $r$ would be Profits and Losses (PnL). The policy would be a Neural Network – a form of artificial intelligence that learns from data which decisions to make. The policy evolves as the agent learns from ongoing feedback, continuously refining the rules and tactics it follows.

RL differs from supervised learning, where the model is trained on a dataset containing input and output with explicit correct answers, as well as from unsupervised learning, which deals with finding patterns in input data without pre-labelled answers. In contrast to the traditional learning methods, RL learns through interactions, progressively improving its policies through continuous feedback and learning. This approach is particularly useful in environments where sequential decision-making is crucial, such as in stock market trading, where the goal is to maximise risk-weighted returns over time. Importantly, the objective for the RL agent is not to predict specific asset market movements or future prices; instead, the agent's goal is to learn how to "play" the stock market effectively – much like learning the rules and strategies of complex games, for example, chess or bridge.

The process of learning in RL involves a methodical trial-and-error approach, where agents continually interact with a given environment. Each interaction, termed an episode, consists of a sequence of actions taken by the agent, during which it observes the consequences of its decisions, receives feedback in the form of rewards and updates its policy accordingly before transitioning to the next state. This iterative process continues until the episode concludes. Through those interaction processes, the agent defines its market trading strategy, enhancing its decision-making ability.

In the RL, the methodologies can broadly be classified into two primary approaches: value-based and policy-based, each with distinct mechanisms and goals. Therefore, value-based methods focus on estimating the value of each action *a* based on current state *s* and choosing the one with the highest estimated value, expressed as *V*(*s*) or *Q*(*s, a*) (Q-Value), which will be explained in detail in chapter 2. These values represent the expected long-term reward of the current policy $\pi$ from the state *s* after taking action *a*. The primary example of this approach is the Double Deep Q-Network (DDQN). Contrarily, policy-based methods directly optimise the policy $\pi$ without computing the estimated value of action *a*. These methods are based on expected values of probabilities, focusing on improving the expected long-term rewards by adjusting the policy parameters directly. Additionally, there is a combination of those two approaches called actor-critic mechanisms. Those combine the best of both worlds, where the actor learns policy with a policy-based approach, while the critic evaluates policy with an estimation of the value function and the values of chosen actions. An example of such an algorithm is Proximal Policy Optimisation (PPO).

Additionally, it is worth noting that all DRL agents presented in this paper are model-free. This means that we do not utilise an explicit model of the market environment for predictions. In other words, agents do not forecast future states of the market, i.e. prices. Instead, these agents learn to make decisions and develop trading strategies based on direct interactions with the market. In this context, the agents are trained to optimise trading actions rather than predict market directions, such as trends or volatility. This model-free approach allows the agents to focus on achieving the best possible trading outcomes based on the reward structures defined, rather than fitting a model to the complex and often non-stationary market dynamics. The decision to use model-free DRL agents is driven by their ability to adaptively learn and improve strategies through trial and error, exploiting profitable opportunities in highly stochastic environments without the need for predefined models.

By interacting with the market, the agent develops a strategy based on the reward system, which typically rewards profits and penalises losses. This method stands in contrast to trying to forecast precise market directions, which is often fraught with uncertainty due to the market's complex and dynamic nature. The agent focuses on maximising its total potential reward by making a series of decisions that best navigate the intricacies and volatilities of the market environment. This strategy allows the agent to adaptively respond to market conditions, optimising trading decisions based on learned experiences rather than predictive assumptions. The model of RL is very close to the behaviours of real investors or day traders who rely on a mix of strategies, intuition, and responsive decision-making to maximise their returns. Just as these traders adjust their strategies based on market performance and conditions, so does the RL agent refine its approach as it learns from the market's responses to its actions. This adaptive, experience-based strategy is key to both human and artificial agents in achieving long-term financial success in the highly variable world of financial trading. In this sense, RL allows artificial intelligence to mirror human learning and decision-making processes more closely than ever before.

## 1.2. Relevance to Efficient Market Hypothesis

The relevance of examining DRL in the context of financial markets extends beyond theoretical interest. It offers a practical framework for understanding how sophisticated algorithms can interact with and potentially capitalise on financial market dynamics. As we delve deeper into algorithms and the implications of DRL within an algorithmic trading framework, it is essential to understand the foundational theory behind financial markets – Efficient Market Hypothesis (EMH), proposed by Fama in his seminal works (1965, 1970, 1992). This thesis was later followed by an enormous range of discussions. On one side, we have papers and books asserting that it is impossible to achieve consistent gains, such as *A Random Walk Down Wall Street,* which argues that stock prices follow a random path and cannot be consistently outperformed (Malkiel, 1973). On the other side, we have arguments supporting the possibility of market inefficiencies and potential gains, such as *A Non-Random Walk Down Wall Street,* which suggests that certain predictable patterns exist (Lo and MacKinlay, 1999). Additionally, *Adaptive Markets: Financial Evolution at the Speed of Thought* proposes that market efficiency evolves over time, allowing for the possibility of gaining an edge in particular time frames (Lo, 2017). These examples illustrate just a few perspectives from the broader discussion spectrum.

The Efficient Market Hypothesis (EMH) assesses that financial markets are 'informationally efficient,' meaning that asset prices reflect all available information at any given time. According to this hypothesis, it is impossible to consistently achieve returns exceeding average market returns on a risk-adjusted basis, given that price changes are driven only by new information that is random and unpredictable. EMH is classified into three forms: 'weak', which means that all past information (like historical prices) is already in the current market price; 'semi-strong', which adds all public information available (like news reports, economic data); and 'strong', which includes all non-public information (i.e. insider information). Therefore, the direct implication of EMH is that it is impossible to consistently

'beat the market' on a risk-adjusted basis because prices always incorporate and reflect all relevant information.

Secondly, employing DRL to test the EMH can offer insights into the potential for outperforming the market through adaptive strategies. If DRL agents can develop trading strategies that consistently yield above-average returns that 'beat the market,' this would directly challenge the EMH. Such results would mean that patterns or inefficiencies exist within the market prices, which could be exploited with the most sophisticated algorithms.

Thirdly, the analysis of DRL agent's behaviour could provide important information about EMH. By examining the strategies and decision-making processes of successful DRL agents, it would be possible to identify the missing piece or specific tactics that enable these agents to outperform benchmarks. This could reveal underlying inefficiencies or exploitable patterns in the market that are not accounted for by the EMH.

**1.3. Literature review**

In this section, we address the lack of literature on deep reinforcement learning (DRL) in algorithmic trading and the typical errors that are common in the available studies in the field of algorithmic trading. Despite the limited number of studies, several methodological concerns frequently arise. Many papers focus on a single asset, which may not provide a comprehensive understanding of the trading landscape. Furthermore, in this case, we could not be sure if individual stock characteristics are responsible for abnormal performance. Additionally, there is often a lack of clear methodologies for cross-validation in time series data, such as the absence of walk-forward optimisation on multiple components. Moreover, some studies are conducted on baskets of stocks that may be prone to survival bias, meaning these studies only include stocks that have not been delisted, potentially questioning the results. Addressing these methodological issues is crucial for developing more reliable and generalisable algorithmic trading strategies.

It is also worth noting that in this highly competitive field, researchers and practitioners avoid sharing their findings, as revealing a successful trading strategy could diminish its effectiveness due to increased adoption. The outcome of this fact is that we end up with a significant gap in the literature that is not found in other fields. The practice of keeping methodologies and results secret contributes to unreliable conclusions and methods revealed in available papers. Therefore, it diminishes the trustworthiness of the available research. Consequently, this paper will not include a broad overview of the literature about algorithmic trading. Instead, we will provide reliable results for DRL with all the necessary components, ensuring that findings are reliable in the field of algorithmic trading.

On the other hand, there is a large range of papers about RL and DRL, which we will briefly discuss. All things presented here will be explained and presented in the next chapter. Reinforcement learning as a formal framework was introduced in the 1980s (Sutton, 1984) and rooted in earlier work on optimal control, as well as decision-making problems. The idea was to develop algorithms capable of learning optimal policies by interacting with an environment, using reward signals as guidance. However, the foundation of these methods was the invention of dynamic programming by Richard Bellman, from which the seminal 'Bellman equation' is derived, serving as the cornerstone for all reinforcement learning methods (Bellman, 1957). The next breakthrough was the invention of Temporal Difference (TD) by Sutton, who is considered one of the fathers of modern RL methods (Sutton, 1984). Next, the State–Action–Reward–State–Action (SARSA) approach was introduced, which focuses on learning policies directly from the experience of sequences of states and actions, which enhances agents by allowing evaluation of policy learning while still interacting with the environment (Rummery and Niranjan, 1994). The Lambda approach to TD, proposed by Sutton and known as TD($\lambda$) (Sutton, 1988), enhances the method of discounting future rewards in the current evaluation of agents. This approach combines elements of the previously proposed temporal difference method with Monte Carlo Reinforcement Learning. Additionally, eligibility traces were introduced as a feature in TD($\lambda$), enabling the method to more effectively utilise experiences from multiple steps back, thus optimising learning from delayed rewards across a sequence of actions.

Addressing the development of methodologies in reinforcement learning (RL), it is crucial to underscore the significant strides made in the ways agents choose actions. The Q-learning algorithm marked the first major value-based method (Watkins; 1989). This approach enables an agent to learn the optimal action-value function, focusing on maximising the expected future rewards from each state via Q-values. Following the development of Q-learning, the field evolved to embrace policy-based methods, which contrast with value-based approaches by directly optimising the policy itself rather than the value of actions (Williams; 1992). This shift led to the adoption of the policy gradient approach, which uses the computation of gradients of expected rewards and uses these gradients to directly adjust the policy. This method is particularly advantageous in environments with continuous action spaces. Further integration of these two approaches resulted in the development of the so-called "Actor-Critic" methods (Konda and Tsitsiklis; 2000). These methods merge the strengths of value-based and policy-based learning: the 'actor' component updates the policy based on the gradients suggested by the 'critic', which evaluates the chosen actions by estimating the value functions.

These methods, along with SARSA, TD and TD(λ), initially required extensive computational resources, making them challenging to implement. The introduction of deep learning, specifically through the use of Neural Networks (NN), marked a pivotal shift. NN, serving as robust approximation tools, allowed for the efficient processing of data. This implementation of DL enables the evolution from a theoretical approach to a practical one, giving birth to modern reinforcement learning—deep reinforcement learning.

The first notable introduction of Deep Reinforcement Learning (DRL) was the development of Deep Q-Network (DQN) by researchers at DeepMind in 2015. DQN combines Deep Neural Networks with the Q-learning algorithm, enabling agents to achieve human-like levels of performance in complex environments, as notably demonstrated on various Atari games (Mnih et al., 2015). This approach allowed the NN to act as a function approximator within the Q-learning framework. Building on the success of DQN, the Double Deep Q-Network (DDQN) was introduced to address the overestimation of action Q-values that DQN can suffer from. DDQN modifies the DQN algorithm by splitting the selection and evaluation of the action in the update step, using two networks to reduce overfitting and improve the stability of the learning process (Hasselt et al., 2016).

Following the foundational work on value-based methods such as DQN, a family of DRL gradient methods, including the actor-critic model known as the Deep Deterministic Policy Gradient (DDPG), was introduced (Lillicrap et al., 2016). Building on and extending on the concept of actor-critic, multiple models were proposed: Asynchronous Advantage Actor-Critic (A3C) (Mnih et al., 2016) and its synchronous counterpart, Advantage Actor-Critic (A2C) (Mnih et al., 2016). Those models optimise the learning process by training with multiple agents in parallel environments. The advancements in DRL have also seen significant contributions from techniques like Generalised Advantage Estimation (GAE) (Schulman et al., 2016). GAE provides an effective approach for estimating the advantage function, which helps in determining how much better an action is compared to the policy's average (alternative actions). This approach is useful with Proximal Policy Optimisation (PPO) (Schulman et al., 2017). PPO simplifies the policy gradient calculations and enhances their performance. Additionally, PPO introduces novel objective functions with policy clips to ensure stability and lower variance in the learning process.

It is important to note that this examination only scratches the surface of the vast field of DRL. This short literature overview only highlights the components that are relevant to our research. For a deeper understanding of RL and DRL, we recommend the comprehensive book by Sutton and Barto (2018). Sutton's study provides an extensive exploration of the foundational theories and practical applications of RL and DRL, making it an invaluable resource for everyone looking to explore these fields further.

# CHAPTER II
# Methodology of Deep Reinforcement Learning

## 2.1. Overview of Deep Learning in DRL

Deep Reinforcement Learning (DRL) combines the learning approach of Reinforced Learning (RL) with Deep Learning (DL) to effectively determine the best action $a$ to choose in a given state $s$ via policy $\pi$. The function approximates those with DL algorithms, primarily Neural Networks (NN). Therefore, this paper focuses on Neural Network and Transformer Network as the key technologies behind the calculations of DRL. In the following subsections, we briefly revisit these foundational concepts.

### 2.1.1. Neural Networks

Neural Networks (NN) are a subset of machine learning algorithms inspired by the neuronal structure of the human brain. These networks are pivotal in deep learning, enabling machines to recognise complex patterns within data. The architecture of a NN typically consists of multiple layers of interconnected nodes (neurons). Each neuron processes input through its activation function and passes the result forward, creating a network capable of learning intricate relationships within data.

Feedforward Neural Networks, also known as multilayer Perceptrons (MLPs), are the simplest type of artificial Neural Network. They are organised into layers: an input layer, one or more hidden layers and an output layer. The input layer receives raw data, which is later processed by hidden layers with weighted connections and activation functions. The output layer is the final output - value or prediction. In the case of our agents, it would be a position taken in the market.

Learning in NN involves adjusting the weights of the connections to minimise the difference between the predicted output and the actual true output. The adjustment is done via various algorithms, with the most popular ones based on gradient descent. The loss function quantifies the errors in predictions. The algorithm that efficiently computes gradients for the loss function across all neurons in the network is called backpropagation. It works by propagating the error backward from the output layer to the input layer, adjusting the weights of each connection to minimise the overall error of the network's predictions.

To further enhance the capabilities of Neural Networks, especially in handling sequential time series data, specialised architectures like Long Short-Term Memory (LSTM) and Gated Recurrent Units (GRU) are employed. These types of NN are designed to capture dependencies and patterns over time, making them suitable for tasks involving time series. LSTM networks address the vanishing gradient problem commonly encountered in NN by introducing memory cells that can maintain their state over long periods of time. An LSTM unit is composed of three gates: the input gate, the forget gate and the output gate. These gates regulate the flow of information into and out of the cell, allowing the network to retain or discard information as needed (Hochreiter, 1997). On the other hand, GRU offers a simpler alternative to LSTMs by combining the forget and input gates into a single update gate. This reduction in complexity leads to fewer parameters, which can result in faster training times while still maintaining the ability to capture long-term dependencies (Cho et al., 2014).

While NN are a critical element of our discussion on DRL, they are a vast topic. Although they are important elements of function approximation in DRL, the main focus of our research is on learning mechanisms, configurations and other related elements in DRL. For more detailed information about Neural Networks, we suggest Aurélien Géron's book (2019).

**2.1.2. Transformer Network**

Transformer networks, first introduced in the seminal paper *Attention is All You Need* (Vaswani et al., 2017), represent a breakthrough in handling sequential data, particularly in the fields of natural language processing and others. These models are characterised by their use of the attention mechanism, which enables them to process input data in parallel and capture complex relationships within the ordered data.

The core innovation inside the Transformer model is the self-attention mechanism, which allows the model to weigh the importance of different elements in data in relation to each other. This mechanism computes a set of attention scores, which dictate how much focus should be placed on other parts of the input sequence when processing a particular element. In addition to the attention mechanism, the Transformer model also utilises positional encodings to retain the order of the input sequence, making them suitable for sequence data i.e. time series input. The architecture of the Transformer consists of two main components: an encoder and a decoder each composed of multiple identical layers. The encoder processes the input sequence and generates a set of continuous representations, while the decoder uses these representations along with the self-attention mechanism to produce the output sequence. Each layer in both the encoder and decoder contains a multi-head self-attention mechanism and a feed-forward NN, followed by layer normalisation and residual connections.

Transformers exhibit several benefits over conventional NN-based models. They eliminate the need for recurrence in model architecture, leading to significant improvements in training speed. The self-attention mechanism allows each position in the encoder to attend over all positions in the previous layer of the encoder, which is particularly effective in modelling relationships and dependencies, regardless of their distance in the input sequence. Additionally, Transformers maintain high scalability and adaptability to various types of data, especially time series one.

For further explanation about Transformers we would like to suggest an original paper by Google DeepMind researchers (Vaswani et al., 2017). It is important to note, however, that the main focus of our study is Deep Reinforcement Learning, while Neural Networks as well as Transformers are the only approximating tool in this branch of machine learning. It is essential to recognise that in any case we would use NN as Neural Networks shortcut, i.e. weights that approximate RL methods, it can be interchangeably understood as referring to the Transformer network as well.

## 2.2. Temporal Difference Learning

Temporal Difference (TD) Learning is a fundamental concept in the field of reinforcement learning (RL). This is what makes RL unique in comparison to supervised and unsupervised learning. TD is a combination of Monte Carlo ideas and Dynamic Programming methods. The main idea is to learn from incomplete episodes, or in other words, to discount future rewards, by bootstrapping them from the current estimates. Monte Carlo methods rely on averaging sample returns to estimate the value of states s or actions a. They require complete episodes to calculate returns, which can make them less efficient in scenarios where continuous learning is needed or where episodes are long or non-terminating. However, Monte Carlo methods provide unbiased estimates, as they rely on actual returns rather than approximations. On the other hand, bootstrapping refers to updating episodes based on other learned estimates without waiting for the final outcome. This means that the value of state is updated using the value of subsequent state. An intuitive explanation of TD would be to not look at only one observation return of position, but to look indefinitely into the future, discounting it with an exponentially-weighted average. By leveraging TD Learning, an RL agent can effectively estimate the long-term benefits of actions taken in the present.

It is essential to highlight that TD learning does not introduce a look-ahead bias, as might be suggested by the leading subscripts of formula [1]. Such bias is avoided because TD learning updates values using observed rewards and current estimates during training, without anticipating or depending on future outcomes. This approach ensures that the learning process remains firmly based on the data available up to the current time step. Moreover, reinforcement learning is applied exclusively to the training dataset, ensuring that no information from validation or test samples is used.

Additionally, crucial to mention is the fact that we would follow Sutton's notation from *Reinforcement Learning: An Introduction* (Sutton, 2018).

### 2.2.1. TD(0) and TD(λ)

TD(0) is the simplest form of Temporal Difference (TD), where we update the value of estimates by the subsequent state, as we look only by one step into the future. Therefore, it is called "(0)", the formula is:

$$V(s_t) \leftarrow V(s_t) + \alpha \left[ r_{t+1} + \gamma V(s_{t+1}) - V(s_t) \right] \qquad [1]$$

where:

$\leftarrow$ - an update formula

$s_t$ - state of the environment at time $t$

$V$ - estimate of the value function at state $s_t$

$t$ - time step

$r_{t+1}$ - reward received after action $a_t$ i.e. after a transaction from $s_t$ to $s_{t+1}$

$\gamma$ - a discount factor, in range [0, 1), this determines how value of future rewards influences the current state

$\alpha$ - learning rate, in the range [0, 1), i.e. how much new information updates the old policy $\pi_{OLD}$

$\left[ r_{t+1} + \gamma V(s_{t+1}) - V(s_t) \right]$ - the Bellman error representing the difference between the current estimate and the new data, reflecting how "off" the current estimate is from what it potentially should be after observing new outcomes

The TD(λ) or TD-lambda algorithm is indeed an extension of the basic Temporal Difference (TD) learning method, which combines the ideas of both TD and Monte Carlo methods. Additionally, it introduces the concept of eligibility traces, which provide a way to bridge the gap between immediate and delayed reinforcement, offering a more flexible mechanism for updating values based on experience.

Update formula in TD(λ) algorithm is given by:

$$V(s_t) \leftarrow V(s_t) + \alpha \sum_{k=0}^{\infty} \gamma^k \lambda^k \delta_{t+k} Z_{t+k} \qquad [2]$$

where:
- λ - represents the decay parameter
- $\sum_{k=0}^{\infty}$ - sum that represents the total sum of decayed future steps, which are used for Value function evaluation
- $\delta_{t+k}$ - Bellman Error at time $t + k$
- $Z_{t+k}$ - eligibility trace at step $t + k$, which gives "credit" to the state $s_t$ for the reward received later

Eligibility traces in the TD(λ) algorithm are a fundamental innovation that allow for a more nuanced approach to learning from sequences of actions, not just single steps. They essentially provide a "memory" of past states, linking the effects of past decisions to future outcomes, much like a story that unfolds over time. To analyse how it works, let's imagine a DRL agent navigating through a series of decisions, such as buying, holding, or selling stocks in a volatile market. Each action the agent takes leaves a trace — an eligibility trace — which captures the significance of past states influenced by the agent's decisions. As the agent moves from $s_t$ to $s_{t+1}$, each time is tagged with eligibility trace $Z_{t+k}$. This trace represents the "credit" that the state $s_t$ should receive for rewards obtained in $t + k$. For example, if we faced a correction in price and the agent took the decision to hold an asset, after which the asset would substantially gain after several *t*, the eligibility traces would help to assign some of that credit for this future back original decision. With this approach, the agent should be able to focus much more on long term investments. In other words, an agent learns from the entire outcome chain that follows its action, which would be totally impossible with classical time series modelling that only focuses on immediate price movements. That's the crucial concept behind TD(λ).

## 2.2.2. Generalised Advantage Estimation

Another groundbreaking concept in RL that takes the idea of TD to the next level is Generalised Advantage Estimation (GAE). GAE is a sophisticated technique that builds on TD learning methods to significantly enhance policy optimisation algorithms, particularly in continuous action spaces. GAE was introduced by Schulman et al., (2016) as a way to effectively balance the bias-variance trade-off in policy gradient estimators, offering a more balanced and efficient way to compute the gradients used to update policies in RL.

Let's start with the answer to the question: what is Advantage Estimation? In the RL, the "advantage" measures the quality of $a_t$ taken in a given $s_t$ over the average action $a$ in that state $s$. Mathematically, the advantage function $A^\pi(s_t, a_t) = Q^\pi(s_t, a_t) - V^\pi(s_t)$, where $Q^\pi(s_t, a_t)$ is the action-value function that estimates the expected return after at in $s_t$ under policy $\pi$, while $V^\pi(s_t)$ is the function value that estimate expected return from $s_t$ under policy $\pi$. So in other words, the advantage function expresses the relative benefit of choosing $a_t$ in comparison to the policy average (alternative actions $a'$).

GAE builds on these basic advantage concepts and aggregates them indefinitely into future discounting estimations of future advantages. This aims to reduce bias and variance. By accurately estimating the advantage, we can better guide the updates to our policies, leading to more stable and faster learning. This is where Generalised Advantage Estimation comes into play.

Therefore, the GAE formula is given by:

$$A_t^{GAE(\gamma,\ \lambda)} = \sum_{k=t}^{T-1}(\gamma\ \lambda)^{k-t}\delta_k \qquad [3]$$

where

$A_t^{GAE(\gamma,\ \lambda)}$ - the Generalised Advantage Estimation at time $t$

$\delta_k$ - TD error at time $k$; calculated as $\delta_k = r_k + \gamma\, V(s_{k+1}) - V(s_k)$ where $r_k$ is reward at time $k$ and $V(s_k)$ is estimated value function at state $s_k$ $(\gamma\ \lambda)^{k-t}$ - discounting and decaying contribution of the estimated TD errors from future actions.

$\sum_{k=t}^{T-1}$ - aggregated sum of all TD errors contributions

Thus, GAE utilises the accumulated sum of discounted TD errors. This creates a more accurate and smoother estimate of the advantage function which is crucial for optimising policies through gradient descent techniques.

Just as TD(λ) takes TD(0) to the next level by considering multiple future steps, GAE leverages the basic advantage estimation to the next level by incorporating discounted future advantages.

## 2.3. Deep Reinforcement Learning algorithms

In the preceding chapter, we introduce the concept of Temporal Difference (TD) learning, detailing how it is refined and extended through various methods such as TD(λ), Advantage, and Generalized Advantage Estimation (GAE). These methods progressively enhance the efficiency and accuracy of value estimation in reinforcement learning, forming a crucial foundation for the more advanced models discussed in this chapter.

In this chapter, we explore Deep Reinforcement Learning (DRL) algorithms, tracing a similar evolutionary path. We begin with the fundamental value-based method, Q-Learning, which, while effective, faces practical limitations in large state spaces. This leads us to the Deep Q-Network (DQN), which integrates deep learning techniques to approximate the Q-function, addressing the scalability issues of traditional Q-Learning. Building on DQN, we delve into the Double Deep Q-Network (DDQN), which mitigates the overestimation bias inherent in DQN by employing two separate networks for action selection and evaluation. This refinement significantly improves the stability and performance of value-based DRL methods.

Next, we transition to the Actor-Critic method, a hybrid approach combining policy-based and value-based strategies. The Actor-Critic framework leverages Q-learning for the Critic's evaluation while the Actor optimizes the policy, using the advantage function instead of TD(λ), thus enhancing learning efficiency. Finally, we introduce Proximal Policy Optimisation (PPO), the most advanced model in this chapter. Additionally, PPO employs the Generalized Advantage Estimation (GAE) and introduces policy clipping to reduce variance in policy updates, ensuring stable and efficient learning. This progression from basic to advanced models illustrates how each successive method builds on the concepts and limitations of its predecessors, culminating in sophisticated algorithms capable of tackling complex reinforcement learning problems.

**2.3.1. Q-Learning**

A Q-Learning algorithm is a model-free, value-based reinforcement learning technique used to find the optimal action-selection policy using a Q-function. Q-learning works by maintaining a table (Q-table) that stores Q-values, which are estimates of the total reward an agent can expect to accumulate over the future, starting from a given state and taking a particular action. The objective of Q-learning is to learn the optimal policy by learning the optimal Q-values for each state-action pair.

Equation for updating with Q-learning:

$$Q(s_t, a_t) \leftarrow Q(s_t, a_t) + \alpha[r_{t+1} + \gamma ( max_{a'} Q(s_{t+1}, a') - Q(s_t, a_t)] \qquad [4]$$

where:
- $\leftarrow$ - an update formula
- $s_t$ - state $a_t$ time $t$
- $Q(s_t, a_t)$ - the current Q-value estimate for being in state $s_t$ and taking action $a_t$
- $\alpha$ - learning rate, which determines how new information should update old policy
- $r_{t+1}$ - a reward received after $a_t$ and moving from $s_t$ to $s_{t+1}$
- $\gamma$ - discount factor
- $max_{a'} Q(s_{t+1}, a')$ - the maximum predicted Q-value for the next state $s_{t+1}$ considering all possible actions $a'$ from $s_{t+1}$

This update formula is derived from the Bellman equation, which forms the theoretical foundation for many dynamic programming and reinforcement learning algorithms. It iteratively adjusts the Q-values towards their true values by using the rewards obtained from the environment as feedback. However, while theoretically Q-learning is powerful it is impractical in environments with large state spaces. In such cases, Q-learning faces many scalable issues. To overcome this, the Deep Q-Network (DQN) employs a deep Neural Network to approximate the Q-function, effectively extending Q-learning to handle larger and more complex environments. This makes DQN an ideal solution to the scalability challenges faced by traditional Q-learning.

**2.3.2. Deep Q-Network (DQN)**

Deep Q-Network (DQN) combines the theoretical approach of Q-learning with the practical usage of Neural Networks (NN), introduced by DeepMind (Mnih et al., 2015). DQN was first demonstrated to achieve human-level performance on many Atari 2600 games using raw pixels as input. The success of DQN demonstrated its potential in other fields such as algorithmic trading. DQN uses a NN to approximate the Q-value function, represented as $Q(s, a; \theta)$ with deep learning. Here, $\theta$ are NN's weights. The approximation of the Q-value equation is given by $Q(s, a; \theta) \approx Q^*(s, a)$. This approach enables the agent to learn the optimal policy through direct interaction with the environment. The update formula for Q-values in DQN is given by:

$$Q(s_t, a_t; \theta) \leftarrow Q(s_t, a_t; \theta) + \alpha[r_{t+1} + \gamma \left( max_{a'} Q(s_{t+1}, a'; \theta^-) \right) - Q(s_t, a_t; \theta)] \quad [5]$$

where:

$Q(s_t, a_t; \theta)$ - the Q-value of the current state-action pair under the current parameters $\theta$.

$\theta$ - weights of Primary Neural Network

$\gamma$ - discount factor

$max_{a'} Q(s_{t+1}, a'; \theta^-)$ - the maximum predicted Q-value for the next state $s_t$ using Target network parameters $\theta^-$

$\theta^-$ - weights of Target Neural Network

The goal of the loss function is to minimise the difference between predicted Q-values and Q-values estimated with TD(λ) via Bellman equation, which is involves the Target network. In the DQN, the Primary neural network learns to evaluate the current state-action pairs and approximate Q-values using TD learning. Meanwhile, the Target network, with is its slower-changing weights, focuses on estimating future states and their discounted rewards, guiding the Primary network's learning process. By minimizing the discrepancy between predicted and target Q-values, the primary network iteratively improves its ability to choose optimal actions and evaluate their corresponding predictions, ultimately refining the agent's decision-making capabilities over time.

Therefore, the loss function for DQN agent is:

$$L(\theta) = E[(r_{t+1} + \gamma \ max_{a'} Q(s', a'; \theta^-) - Q(s_t, a_t; \theta))^2] \qquad [6]$$

where:

$max_{a'} Q(s', a'; \theta^-)$ - maximum predicted Q-value for $s_{t+1}$, considering all possible $a'$. This is calculated with the Target network $\theta^-$

The target network guides the learning of DQN agent to prioritize actions in a given state that lead to the highest discounted future rewards. Additionally, DQN incorporates experience replay, which stores a batch of experience agent got from playing the stock market. Hence, the agent would be learning and updating his network weights after the full batch was collected (we won't be updating network weights after each step, but rather after n steps), which is a standard approach in deep machine learning.

### 2.3.3. Double Deep Q-Network (DDQN)

The Double Deep Q-Network (DDQN) builds on the foundation of the standard Deep Q-Network (DQN) by addressing a critical issue identified in DQN: the tendency to overestimate Q-values (Hasselt et al., 2016). This overestimation is due to a mechanism that updates Q-values by taking the max operator ($max_{a'} Q(s', a'; \theta^-)$). As we use the max of the next state to update Q estimates and continue to take the max in subsequent steps (*t+2*, *t+3*, etc.), this compounding max operation leads to biassed and overly optimistic Q-values. The key modification in DDQN lies in its approach to updating Q-values estimates. Instead of using a single network for both selecting and evaluating actions, DDQN employs two separate networks. In DDQN, the Primary network is used for selecting the best action based on the highest Q-value and the Target network is used for evaluating the Q-value of that selected action. In the calculation of target Q-value estimates with the Target network, the primary network is also used to determine the action to be evaluated (hence "double" in the name). This method involves separate roles for action selection and evaluation of Q-value estimates, reducing overestimation bias from the vanilla version (DQN).

Updating Q-values mechanism in DDQN can be described by:

$$Q(s_t, a_t; \theta) \leftarrow Q(s_t, a_t; \theta) + \alpha[r_{t+1} + \gamma Q(s', \arg max_{a'} Q(s', a'; \theta); \theta^-) - Q(s_t, a_t; \theta)] \quad [7]$$

where:

$\leftarrow$ - an update formula

$\theta$ - weights of Primary Neural Network, chooses best $a$

$\theta^-$ - weights of Target Neural Network, evaluate chosen $a$

$\arg max_{a'} Q(s', a'; \theta)$ - uses Primary Neural Network to select $a$

### 2.3.4. Actor-Critic

Actor-Critic methods represent a powerful class of algorithms in Deep Reinforcement Learning (DRL) that merge the benefits of both policy-based and value-based approaches. Actor-Critic methods operate on a dual-structure framework. The "actor" is a model that directly determines the policy by suggesting the best possible action to take given the current state of the environment. The representation of this is $\pi(a|s; \theta)$, where, like in previous methods, $a$ is action, $s$ is current state and $\theta$ are parameters defining the policy (weights of actor's NN). On the other hand, the 'critic' evaluates the action chosen by the actor. Critic estimates the value function $V(s, \omega)$, where $\omega$ are the parameters of the critic model (weights of critic's NN). Worth to point out is that Critic's evaluation is grounded in Q-learning, often expressed as $Q(s, a; \omega)$.

The Critic aims to evaluate how good actions taken by actors truly are. Those are done using TD learning (TD($\lambda$)) or in our case advantage. The critic updates his NN's weights $\omega$ with loss function given by:

$$L(\omega) = E[(r_{t+1} + \gamma V(s_{t+1}, \omega) - V(s_t, \omega))^2] \quad [8]$$

where:

$\omega$ - critic's NN weights

$r$ - reward

$\gamma$ - discount factor

$s$ – state

Subsequently, the Actor is updated via policy π gradient method, which aims to maximise expected return. The objective function $J(\theta)$ to maximise the expected return is given by:

$$\nabla J(\theta) = E[\log \pi(a|\, s_t;\, \theta)\, A(s_t, a\,;\omega)] \quad\quad [9]$$

where:

$\pi(a_t|\, s_t;\, \theta)$ - probability of taking $a$ at $s_t$, calculated by actor's policy π

$A(s_t, a\,;\omega)$ – advantage of $a$ at $s_t$, which measures the relative value of taking $a$ over alternatives at $s_t$, calculated by critic

By updating both the Actor and Critic iteratively, Actor-Critic methods effectively learn both policy that maximises rewards and value function that accurately predicts future rewards.

The interaction between the policy π (Actor), Value function $V$ (Critic) and the environment is depicted in Figure 1. The Actor's π selects an $a$ based on the $s_t$. The Value function (Critic) evaluates this action $a$ and provides feedback in the form of a TD error, which in our case is represented by the advantage function $A^\pi(s_t, a_t)$. The environment then responds to the action $a_t$ by providing a new state $s_{t+1}$ and reward $r_{t+1}$, which are used to further update the Actor ($\theta$) and Critic ($\omega$).

Figure 1. Actor-Critic method in Reinforcement Learning: Interaction between Policy, Value Function and Environment.

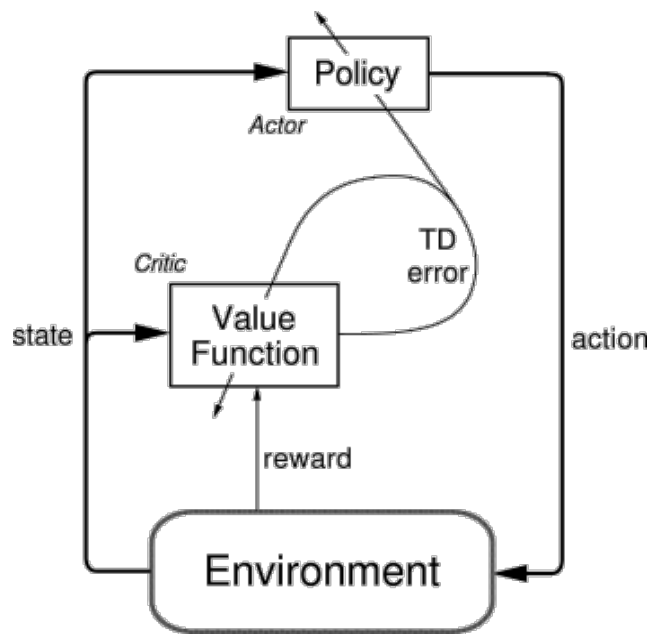

*Source: Sutton, R. S., & Barto, A. G. „Actor-Critic method". Reinforcement Learning: An Introduction. (2018)*

### 2.3.5. Proximal Policy Optimisation (PPO)

Proximal Policy Optimisation (PPO) is a policy-based model that utilizes the actor-critic methodology, combining both value function approximation and policy optimisation. PPO is designed to maintain learning stability and improve the efficiency of policy updates in reinforcement learning (RL) by employing a conservative update strategy. This strategy, known as "clipping," restricts the amount by which the new policy $\pi$ can deviate from the old policy $\pi_{OLD}$ during updates. This clipping mechanism limits potentially harmful large updates, which can destabilize the learning process.

The core of PPO's clipping is to modify standard policy gradient with addition of

$$L^{CLIP}(\theta) = E[\min{(R_t(\theta) A_t^{GAE}, CLIP(R_t(\theta), 1 - \varepsilon, 1 + \varepsilon) A_t^{GAE})}] \qquad [10]$$

where:

$R_t(\theta) = \frac{\pi(a|s_t;\theta)}{\pi_{OLD}(a|s_t;\theta)}$ - probability ratio of the action under current policy $\pi$ to previous policy $\pi_{OLD}$ (which is previous agent generation and previous NN's weights $\theta$)

*CLIP* - represents clipping i.e. preventing big updates in policy $\pi$ (change from $\pi_{OLD}$ to $\pi$)

$\varepsilon$ - small value that sets range of clipping, most commonly are used values in range (0.1 up to 0.25). It ensures that the new policy $\pi$ does not deviate too much from the old policy $\pi_{OLD}$ by limiting the update magnitude

$A^{GAE}$ - Generalised Advantage Estimation. See equation [3]

The clipping mechanism in PPO plays a crucial role in stabilizing the training process. The idea is to prevent new policy $\pi$ for being too different from previous old policy $\pi_{OLD}$, by limiting the magnitude of changes. This approach prevents overly aggressive updates that can destabilise learning process, resulting in the end with poor performance. The clipping operation restricts the probability ratio $R_t(\theta)$ within the range $[1 - \varepsilon, 1 + \varepsilon]$, where $\varepsilon$ is DRL hyperparameter, which is arbitrarily chosen. In the case that $R_t(\theta)$ would fall outside this range, the update would be 'clipped' to stay within the acceptable range. This approach highly helps in maintaining stable learning dynamics. However, the clipping parameter $\varepsilon$ cannot be too small, as this would excessively restrict policy updates, preventing the policy from effectively adapting and learning. On the other hand, if $\varepsilon$ is too large, the clipping mechanism would rarely be used. Balancing the clipping parameter is crucial to ensure that the updates are neither too conservative nor too aggressive.

Moreover, PPO uses GAE instead of Advantage in formula for objective function, to further enhance Actor-Critic method:

$$\nabla J(\theta) = E[log \, \pi(a|\, s_t;\, \theta) \; A_t^{GAE}(s_t, a\,;\omega)] \qquad [11]$$

By integrating these mechanisms, PPO yields a highly effective DRL agent with reduced variance in estimates. It merges the strengths of both value and policy-based approaches through its actor-critic architecture, enhancing the agent's ability to estimate future rewards accurately and adjust its policy based on reliable feedback. Additionally, it uses GAE instead of Advantage.

# CHAPTER III
## Framework of our research

Before we examine the findings of our analysis, it is essential to focus on the practical methodology of our research and the metrics evaluation platform we used to determine the significance of our results.

### 3.1. Research Methodology

In our research, we analyse data from five distinct assets: three of the most traded currencies — Euro (EUR), United States Dollar (USD) and Japanese Yen (JPY), resulting in three currency pairs (EUR/JPY, EUR/USD, USD/JPY); the world's most significant index, the Standard and Poor's 500 (S&P 500); and the most popular cryptocurrency – Bitcoin (BTC). These assets were chosen because they are among the most liquid and widely traded in their respective categories. The selection is somewhat arbitrary but focused on ensuring high liquidity and popularity. Additionally, potential different behaviour of those assets might be expected, especialy in terms of volatility. After each change in position we would reinvest all of the capital into buying or selling. Reinvesting is not taking place when position is not changed. Our trading strategy involved testing these assets on a daily frame, where agents executed actions at midnight US time (GMT -4 hours) every 24 hours.

The input for each model varied depending on the asset but generally included the current market position (to teach the agent about provisioning) and the returns based on the closing prices. For some models, we also incorporated OHLC (Open-High-Low-Close data), technical indicators such as the Relative Strength Index (RSI), moving averages, Average True Range (ATR) and the moving average convergence/divergence indicator (MACD). To address seasonality or time-related trends, we used the sine and cosine of time. All data inputs were standardised or normalised, with careful measures taken to prevent look-ahead bias.

In our study, to train our DRL models, we employ a moving forward optimisation, also referred to as rolling window optimisation, which is depicted in Figure 2. This approach involves dividing the data into multiple sequential windows. In our case, we organise the data into five windows, where each validation set encompasses one year of data and the subsequent out-of-sample test also spans one year. The first validation period is set in 2018 with 2019 designated as the out-of-sample testing year. The following testing periods then progress annually through 2020, 2021, 2022 and 2023. To ensure consistency and relevance of the data to the specific markets we are analysing, the training samples have a fixed start date which varies depending on the asset. For currencies and the S&P 500, data collection begins from the year 2005. For Bitcoin, which entered the financial markets later, the data collection starts from 01/01/2013. With each new window, we initiate training of the agent from scratch. This strategy of reinitializing the learning process for each window prevents any carryover of biases or overfitting from previous training phases to the current learning environment.

Figure 2. Walk-forward optimisation process.

| Iteration\Years | 2005-2016 | 2017 | 2018 | 2019 | 2020 | 2021 | 2022 | 2023 |
|---|---|---|---|---|---|---|---|---|
| First optimisation | train | train | validation | test | | | | |
| Second optimisation | train | train | train | validation | test | | | |
| Third optimisation | train | train | train | train | validation | test | | |
| Fourth optimisation | train | train | train | train | train | validation | test | |
| Fifth final optimisation | train | train | train | train | train | train | validation | test |

*Source: Own study. The table outlines the distribution of training, validation, and testing periods over different years for multiple optimisation iterations.*

A critical component in RL is reward system. In our approach, we experimented with various reward structures. Initially, we tested higher penalties for incorrect actions. However, this approach did not yield better results, as the agents tended to avoid taking any actions, effectively refusing to "play" in long run. In the end we adopted simple Profit and Losses (PnL), where reward was how many % agent got for position. PnL is the most straightforward and interpretable measure. However, without any adjustment agents struggle to differentiate between actions because the returns were too close to zero. To overcome this issue, we included scaling of those rewards by factor (different for different assets). Mostly 100 or 1000. By scaling the PnL reward, we provided clearer incentives for the agents, leading to more effective training outcomes.

In developing our methodology, we select agent's generation that have undergone at least half of their training period. This approach helps ensure that the Neural network weights are appropriately adjusted but not overfitted. Our primary performance metric is the Sharpe Ratio, which we will discuss in more detail in the following subchapter. To verify the robustness of our results, we analyse agents from two generations before and after the chosen generation. This step is crucial to confirm that the observed performance on the validation set is not merely a random occurrence but is consistent across nearby agent's generations (policy $\pi_{OLD}$ and $\pi_{NEXT}$). The generation of the agent is a marker that increments with each training cycle, allowing us to track and compare performance over time.

To prevent overfitting in Deep Reinforcement Learning (DRL), which is a common challenge in the field of Deep Learning, due to the complex nature of the environments and models involved, several strategies were introduced. L1 and L2 regularisation, also known as Lasso and Ridge regularisation, are employed to reduce the model's complexity by penalising the weights of the network. L1 regularisation adds a penalty equivalent to the absolute value of the magnitude of coefficients, promoting sparsity (Tibshirani, 1996), while L2 adds a penalty equal to the square of the magnitude of coefficients, which helps control the model's variance by keeping the coefficients small (Hoerl et al., 1970). Another very effective method to prevent overfitting is using dropout layers. This technique involves randomly deactivating a fraction of the neurons within the neural network at each update during the training phase. Dropout prevents co-adaptation of neurons by making the training process noisy, thereby forcing the neurons to operate independently and making the model more robust. This randomness helps the network to generalise better (Hinton et al., 2013).

Last but not least, very common practice in DRL is to rather than learning on the entire training dataset at once is to cycle through different subsets of training data. This approach allows agent each time to engage with varied data segments during the training cycles, enabling it to encounter a diverse range of scenarios. Each cycle offers the agent new opportunities to select actions, which in turn lead to different outcomes and rewards. This method helps ensure that the agent does not simply memorise specific data patterns but learns to adapt and respond to new situations, enhancing its ability to generalise across different environments. This strategy is not only crucial for achieving robustness but also for ensuring that the agent's performance is reliable and effective outside the controlled training environment.

### 3.2. Evaluation

In this section, we will explore the methods employed to assess the effectiveness of our deep reinforcement learning (DRL) agents in the financial markets. The first natural metric for evaluating our strategy is profitability. Starting with an initial capital of $10,000, it is crucial to decide whether our strategy can achieve a net positive return, taking into account factors such as commissions, bid-ask spread and provisions. However, just profitability does not suffice as a lazy strategy could potentially yield higher returns simply through passive buy-and-hold approach. Therefore, our strategy must not only be profitable but also be capable of outperforming this common strategy. To further challenge our DRL agents, we set a higher benchmark: beating the performance of a "perfect" annual strategy. This benchmark represents an idealised strategy that makes an optimal prediction at the start of each year and holds that position for the entire year. For instance, if the EUR/USD return in 2019 was -2.1%, the "perfect" strategy would go short for the entire year. Conversely, if the EUR/USD return in 2020 was +8.86%, the perfect strategy would switch to long for the whole next year. This benchmark adjusts its position only once a year, based on perfect foresight of the annual return.

Additionally, it is crucial to consider the varying levels of volatility across different financial asset classes. For instance, a daily 3% movement in major currency pairs like EUR/USD is very uncommon, whereas the same percentage change in Bitcoin might indicate a relatively stable day. To accommodate these differences in market behaviour, we can employ metrics such as annualised Sharpe ratio and the Sortino ratio, which help adjust returns based on risk and volatility. The Sharpe ratio measures the performance of an investment compared to a risk-free asset, after adjusting for its risk (Sharpe, 1966) (Equation [12]). It is calculated by subtracting the risk-free rate from the return of the investment and then dividing by the investment's standard deviation of returns. The Sortino ratio, on the other hand, differentiates itself by only considering the downside risk, which is more relevant to investors who are concerned primarily with the negative variance in their returns (Sortino, 1994) (Equation [13]). The Sortino ratio improves upon the Sharpe ratio by distinguishing harmful volatility from total overall volatility. It is worth pointing out that we consider the risk-free return as 0 for this research because it was effectively zero until early 2022. With this assumption, we end up with a measure similar to the information ratio. The annualising factor on daily returns would be the number of days that can be traded in a given year: for currency pairs it would be 6 days per week, for S&P 500 5 days and for Bitcoin 7 days.

Annualised Sharpe ratio Equation:

$$Sharpe\ ratio\ =\ \frac{R-R_f}{\sigma}\sqrt{N} = \frac{R}{\sigma}\sqrt{N} \qquad [12]$$

where

$R$ - return of strategy

$R_f$ - risk free rate in our case equal to 0

$\sigma$ - standard deviation of strategy

$N$ - annualization factor, number of intervals in the year

Annualised Sortino ratio Equation:

$$Sortino\ ratio\ =\ \frac{R-R_f}{\sigma_d}\sqrt{N} = \frac{R}{\sigma_d}\sqrt{N} \qquad [13]$$

where:

$\sigma_d$ - standard deviation of the negative returns only

Moreover, we will evaluate the annualized return using the Compound Annual Growth Rate (CAGR), which provides a smoothed annual rate of return, mitigating the effects of volatility and offering a clearer picture of the investment's performance over time. The CAGR is calculated using the formula:

$$CAGR = \left(\frac{Ending\ Value\ of\ strategy}{Beginning\ Value=10\ 000}\right)^{\frac{1}{Number\ of\ years}} - 1 \qquad [14]$$

Additional metrics such as maximum drawdown and the duration of maximum drawdown can also provide insight into the potential risks associated with the investment strategy. Maximum drawdown measures the largest single drop from peak to bottom in the investment portfolio before a new peak is achieved, and its duration is the time taken to recover from this drop. These metrics are critical in understanding the kind of temporary losses the strategy might incur during a market downturn and how swiftly it can recover.

In addition to the quantitative metrics discussed previously, analysing the behavioural patterns of our DRL agents provides valuable insights into their operational dynamics within the stock market. Observing the agent's decisions, can lead to important conclusions. Factors that we would consider would include:

- **Transactions costs**: how much the agent would spend on provisions (transaction costs, including spreads and commissions). Higher costs could erode profits and make a profitable agent not so.
- **Trading Frequency**: The number of trades executed by the agents over a given period highlights the strategy's activity level. A high number of trades could suggest a more aggressive strategy, which might increase costs and risks, whereas a lower number could indicate a more conservative or opportunistic approach. By number of trades, it is meant number of opened positions.
- **Average Position Duration**: Analysing how long positions are held on average can tell us about the strategy's market exposure and risk appetite. Shorter holding periods might align with a high-frequency trading strategy focusing on arbitrage opportunities, whereas longer durations could suggest a trend-following or buy-and-hold strategy.
- **Position Distribution**: We will also look at the percentage of time the agent spends in different positions during the out-of-sample test period. For example, if the agent was 50% of the time in a long position, 25% in a short position, and 25% without any position, it would indicate the agent's preferred market stance and its flexibility in adjusting to market conditions.
- **Win rate**: The win rate, or the percentage of profitable trades out of the total trades executed, is another crucial metric. A higher win rate might indicate a more reliable strategy, but it must be considered alongside other metrics such as profit factor to ensure the overall profitability and sustainability of the strategy.

By obtaining those behavioural and quantitative results, we will have a full view on DRL performance in algorithmic trading.

# CHAPTER IV
## Results

In this section, we delve into the core of our study: the application of Deep Reinforcement Learning (DRL) in algorithmic trading. This chapter systematically examines the performance of DRL strategies across different asset. The notatios are as follows: DDQN is Double Deep Q-Network and PPO is Proximal Policy Optimisation. NN means classical fully connected Neural Network while T means Transformer Network. These notations help distinguish the various strategies and neural network architectures employed in our study.

On the technical side, provision for major FOREX currency pairs was set at 0.01%, which is extremely low due to liquidity of the market. We haven't included any swap points as well as leverage was not included. In comparison, provision for S&P 500 index was set at 0.025% while for Bitcoin at 0.1%. For the gamma parameter, crucial for TD and GAE for all assets, it was set at 0.75 which would mean average discounting of future. For the length of state (input) the agent was looking backward was set at 20 days, providing a balance between recent and historical data.

Given the substantial computational power required, training each model took approximately one day per asset. Due to these constraints, we did not implement optimization techniques such as random search or grid search. Instead, we relied on the expert knowledge of the authors to select and test parameters. Additionally, short positions were enabled in our trading strategies, allowing the agents to capitalize on both rising and falling markets.

Before presenting the results, it is essential to outline our expectations. Based on existing literature and theoretical foundations, we anticipated that PPO should outperform DDQN due to its robustness and ability to handle more complex environments with its through its Actor-Critic architecture and the use of Generalized Advantage Estimation (GAE) instead of TD($\lambda$). Furthermore, it is expected that the Transformer Network (T) would outperform the classical Neural Network (NN) due to its superior ability to capture long-range dependencies and model sequential data effectively.

The analysis of results is presented in a focused manner, emphasizing key aspects rather than all numerical details. Full results, including detailed numerical data, are provided in the corresponding tables for each asset. Additionally, two key plots, balance over time and drawdowns, are included to visually demonstrate the performance and risk profiles of the strategies.

It is important to note that the graphs in our analysis start from the first trading day i.e. after PnL from first day. Consequently, some curves do not begin at the initial capital of 10,000 units, reflecting the gains from first trading day of 2019. Those which starts indicate first agent's decision to stay out of the market.

### 4.1. EUR-USD

We begin the presentation of our results with the most traded currency pair on the Foreign Exchange Market (FOREX) - EUR/USD. According to Bank for International Settlements (2019), the EUR/USD currency pair accounted for approximately 24.0% of all daily traded volumes in the FOREX market. This makes it the most liquid and widely traded currency pair globally. The EUR/USD exchange rate indicates how many U.S. dollars are needed to purchase one euro. In Table 1, presented are performance results of our study while Figure 3 shows Profit and Losses i.e. total balance of strategy over time and Figure 4 represents maximum drawdown over time.

Table 1. Performance measures for DDQN-NN, DDQN-T, PPO-NN, PPO-T and two benchmarks over 2019-2023 years for the EUR/USD.

| EURUSD | Final balance | Provision Sum | Total trades | CAGR | Annualised std. | Sharpe ratio | Sortino Ratio | Maximum Drawdown | Maximum Drawdown Duration | Average Position Duration | Win Rate | In Long | In Short | Out of the market |
|---|---|---|---|---|---|---|---|---|---|---|---|---|---|---|
| **DDQN_NN** | 14 444.74 | -351.9 | 300 | 7.63% | 4.14% | 1.842 | 2.043 | -4.40% | 167 | 2.28 | 58.6% | 17.7% | 26.2% | 56% |
| **DDQN_T** | 12 512.91 | -227.4 | 202 | 4.58% | 4.43% | 1.035 | 0.853 | -11.30% | 392 | 2.56 | 55.4% | 2.5% | 30.7% | 66.7% |
| **PPO_NN** | 10 360.83 | -522.7 | 506 | 0.71% | 6.53% | 0.108 | 0.136 | -14.70% | 703 | 2.45 | 50.3% | 49.5% | 30.1% | 20.3% |
| **PPO_T** | 13 271.26 | -657.2 | 546 | 5.82% | 6.86% | 0.847 | 1.197 | -11.60% | 394 | 2.45 | 48.5% | 40.4% | 45.6% | 13.9% |
| **Benchmarks:** | | | | | | | | | | | | | | |
| **Buy and hold** | 9 641.81 | -1 | 1 | -0.84% | 7.84% | -0.108 | -0.159 | -22.50% | 931 | 1557 | 100% | 100% | 0% | 0% |
| **"perfect" annual strategy** | 13 150.82 | -4.4 | 4 | 6.61% | 7.07% | 0.935 | 1.396 | -9.20% | 392 | 389.25 | 100% | 40% | 60% | 0% |

*Source: Own study on the 2019-2023 daily data for EUR/USD (1557 observations) with starting capital of 10 000. DDQN is Double Deep Q-Network, PPO is Proximal Policy Optimisation, NN- fully connected Neural Netowork and T – transformer Network. The agents can take position with whole capital between long, short and out of the market. 'Perfect' annual strategy benchmark is a hypothetical trading strategy that predicts the optimal market position at the start of each year and holds that position, for the entire year based on the annual return forecast with perfect foresight.*

Figure 3. Balance changes for DDQN-NN, DDQN-T, PPO-NN, PPO-T and two benchmarks over 2019-2023 years for the EUR/USD.

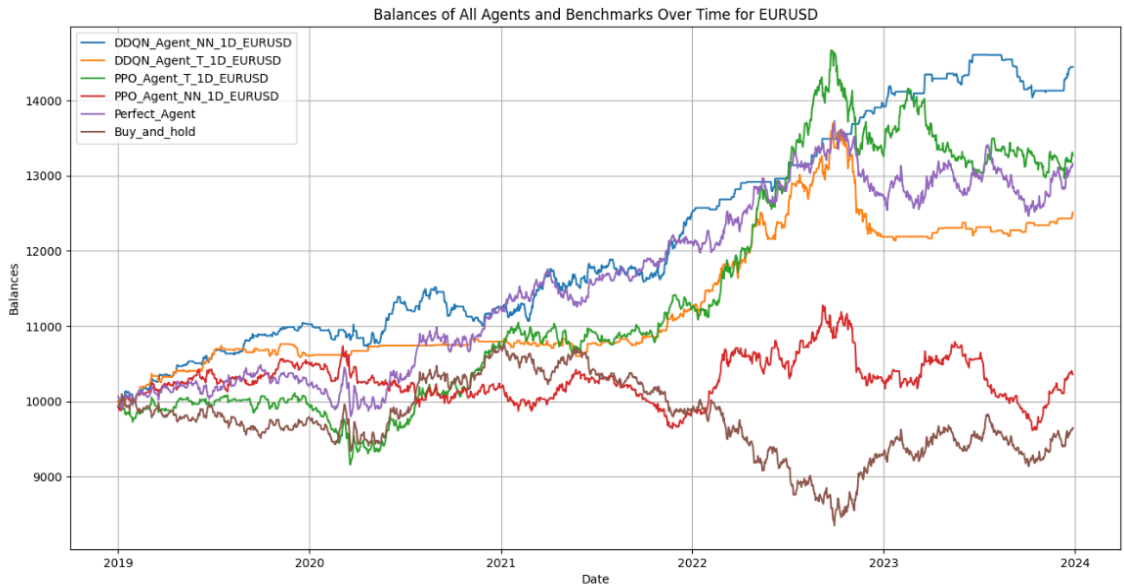

*Source: Own study on the 2019-2023 daily data for EUR/USD (1557 observations) with starting capital of 10 000. DDQN is Double Deep Q-Network, PPO is Proximal Policy Optimisation, NN- fully connected Neural Netowork and T – transformer Network. The agents can take position with whole capital between long, short and out of the market. 'Perfect' annual strategy benchmark is a hypothetical trading strategy that predicts the optimal market position at the start of each year and holds that position, for the entire year based on the annual return forecast with perfect foresight.*

Figure 4. Drawdowns for DDQN-NN, DDQN-T, PPO-NN, PPO-T and two benchmarks over 2019-2023 years for the EUR/USD.

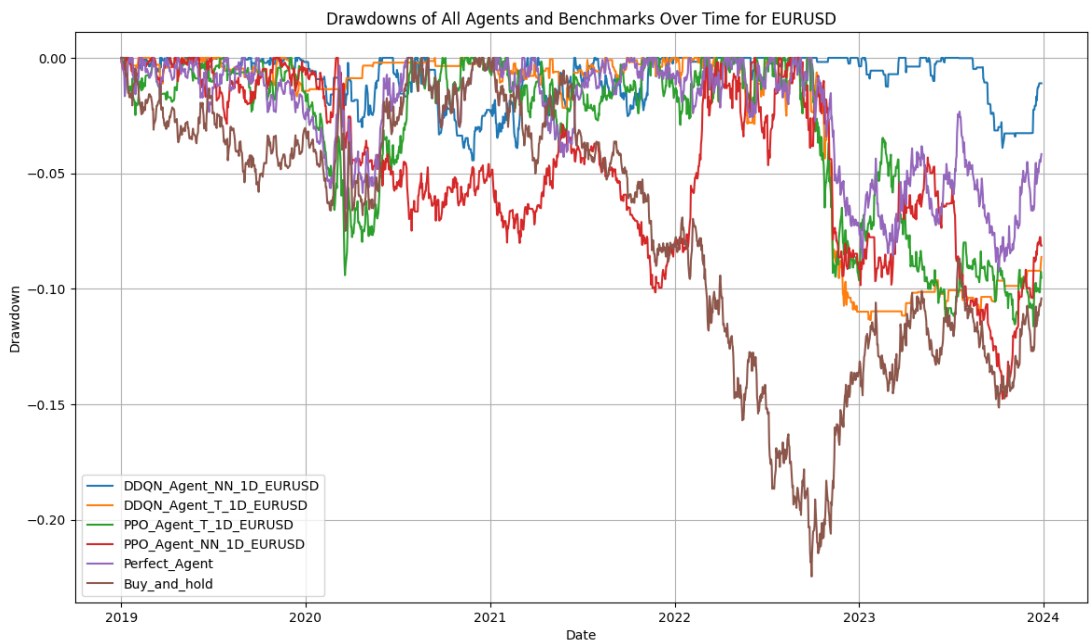

*Source: Own study on the 2019-2023 daily data for EUR/USD (1557 observations) with starting capital of 10 000. DDQN is Double Deep Q-Network, PPO is Proximal Policy Optimisation, NN- fully connected Neural Netowork and T – transformer Network. The agents can take position with whole capital between long, short and out of the market. 'Perfect' annual strategy benchmark is a hypothetical trading strategy that predicts the optimal market position at the start of each year and holds that position, for the entire year based on the annual return forecast with perfect foresight.*

The analysis of various trading strategies on EURUSD over five years, as depicted in the provided Table 1 and Figures 3 and 4, offers critical insights into their performance and key findings. The DDQN_NN strategy emerged as the best performing with a final balance of 14,444.74. This strategy showed strong risk-adjusted returns, outperforming the perfect benchmark strategy. One of the key factors contributing to its success was its conservative approach, with DDQN_NN maintaining a high percentage of time out of the market (56%), thereby avoiding unfavourable market conditions.

Following closely was the PPO_T strategy, which achieved a final balance of 13,271.26. PPO_T also managed to barley surpass the perfect benchmark strategy, although it took on more risk, as evidenced by its high annualized standard deviation of 6.86% and a lower Sharpe ratio of 0.847. This strategy's frequent trading activity, with only 13.9% of the time spent out of the market, contributed to its higher risk profile, resulting in less favourable risk-adjusted returns compared to DDQN_NN.

In terms of risk-adjusted performance, both DDQN_T (Sharpe 1.035, Sortino 0.853) and DDQN_NN (Sharpe 1.842, Sortino 2.043) outperformed the perfect benchmark strategy, indicating higher returns per unit of risk taken. Additionally, DDQN_T and PPO_T performed very well in 2022, capitalizing on market conditions during that year. In contrast, DDQN_NN demonstrated good performance consistently across all years analysed.

Market activity analysis revealed that agents tended to stay out of the market a significant portion of the time, particularly DDQN_T (66.7%) and DDQN_NN (56%). This conservative trading behaviour likely contributed to their superior performance by avoiding periods of high volatility and adverse market movements. On the other hand, PPO_T's high frequency of trades and minimal out-of-market time (13.9%) resulted in the highest annualized standard deviation among the strategies, impacting its risk-adjusted performance.

**4.2. EUR-JPY**

Next the presents are our results for another FOREX currency pair - EUR/JPY. According to Bank for International Settlements (2019), the EUR/JPY currency pair accounted for approximately 4% of all daily traded volumes in the FOREX market. The EUR/JPY exchange rate indicates how many Japanese Yen are needed to purchase one euro. In Table 2, presented are performance results of our study while Figure 4 shows Profit and Losses i.e. total balance of strategy over time and Figure 5 represents maximum drawdown over time.

Table 2. Performance measures for DDQN-NN, DDQN-T, PPO-NN, PPO-T and two benchmarks over 2019-2023 years for the EUR/JPY.

| EURJPY | Final balance | Provision Sum | Total trades | CAGR | Annualised std. | Sharpe ratio | Sortino Ratio | Maximum Drawdown | Maximum Drawdown Duration | Average Position Duration | Win Rate | In Long | In Short | Out of the market |
|---|---|---|---|---|---|---|---|---|---|---|---|---|---|---|
| **DDQN_NN** | 11 960.99 | -412.27 | 396 | 3.64% | 6.57% | 0.554 | 0.639 | -9.44% | 530 | 2.5 | 51% | 49.4% | 14.1% | 36.4% |
| **DDQN_T** | 16 105.72 | -316.08 | 257 | 10% | 7.98% | 1.252 | 1.714 | -8.37% | 233 | 5.45 | 58.7% | 53.6% | 36.3% | 10% |
| **PPO_NN** | 13 980.63 | -309.41 | 273 | 6.93% | 6.17% | 1.122 | 1.198 | -5.76% | 377 | 3.08 | 58.9% | 24.8% | 29.3% | 45.9% |
| **PPO_T** | 18 120.64 | -487.69 | 362 | 12.62% | 7.2% | 1.752 | 2.212 | -9.16% | 324 | 3.01 | 55.5% | 31.1% | 39% | 29.8% |
| **Benchmarks:** | | | | | | | | | | | | | | |
| **Buy and hold** | 12 428.34 | -1.0 | 1 | 5.21% | 9.06% | 0.575 | 0.775 | -9.86% | 577 | 1557 | 100% | 100% | 0% | 0% |
| **"perfect" annual strategy** | 13 340.61 | -2.04 | 2 | 6.97% | 9% | 0.774 | 1.084 | -8.27% | 289 | 778.5 | 100% | 80% | 20% | 0% |

*Source: Own study on the 2019-2023 daily data for EUR/JPY (1557 observations) with starting capital of 10 000. DDQN is Double Deep Q-Network, PPO is Proximal Policy Optimisation, NN- fully connected Neural Netowork and T – transformer Network. The agents can take position with whole capital between long, short and out of the market. 'Perfect' annual strategy benchmark is a hypothetical trading strategy that predicts the optimal market position at the start of each year and holds that position, for the entire year based on the annual return forecast with perfect foresight.*

Figure 5. Balance changes for DDQN-NN, DDQN-T, PPO-NN, PPO-T and two benchmarks over 2019-2023 years for the EUR/JPY.

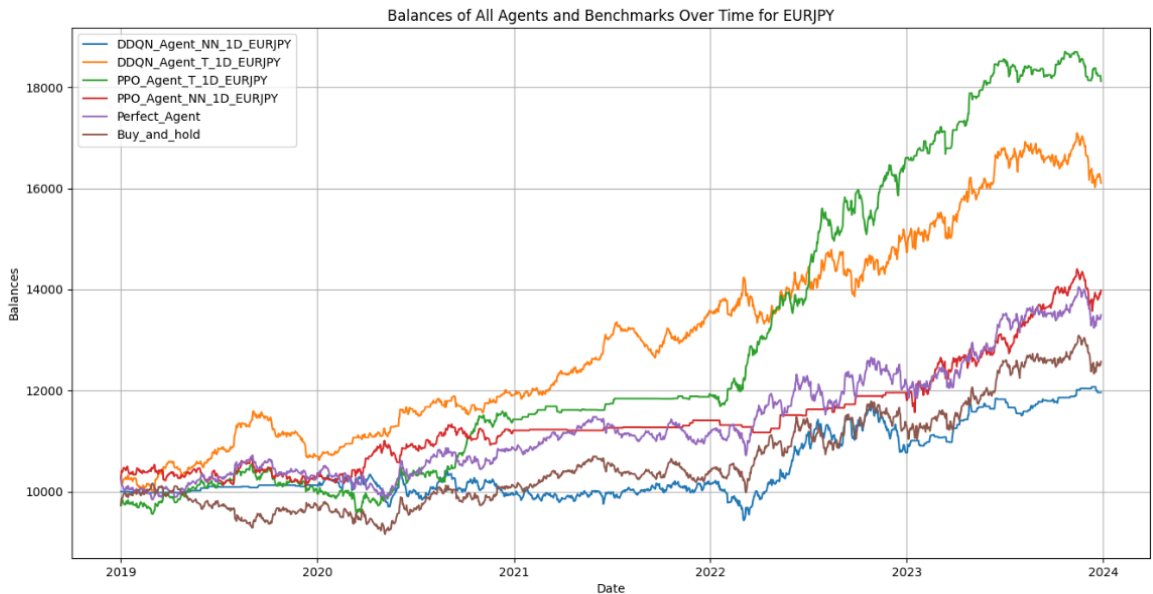

*Source: Own study on the 2019-2023 daily data for EUR/ JPY (1557 observations) with starting capital of 10 000. DDQN is Double Deep Q-Network, PPO is Proximal Policy Optimisation, NN- fully connected Neural Netowork and T – transformer Network. The agents can take position with whole capital between long, short and out of the market. 'Perfect' annual strategy benchmark is a hypothetical trading strategy that predicts the optimal market position at the start of each year and holds that position, for the entire year based on the annual return forecast with perfect foresight.*

Figure 6. Drawdowns for DDQN-NN, DDQN-T, PPO-NN, PPO-T and two benchmarks over 2019-2023 years for the EUR/JPY.

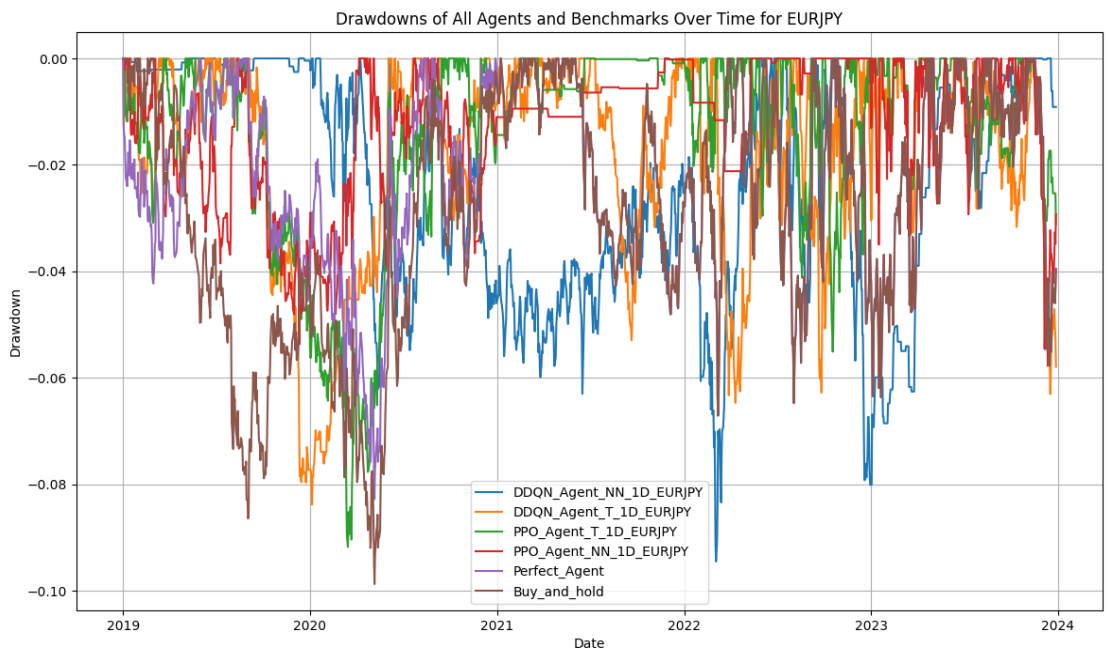

*Source: Own study on the 2019-2023 daily data for EUR/ JPY (1557 observations) with starting capital of 10 000. DDQN is Double Deep Q-Network, PPO is Proximal Policy Optimisation, NN- fully connected Neural Netowork and T – transformer Network. The agents can take position with whole capital between long, short and out of the market. 'Perfect' annual strategy benchmark is a hypothetical trading strategy that predicts the optimal market position at the start of each year and holds that position, for the entire year based on the annual return forecast with perfect foresight.*

The evaluation of trading strategies on the EUR/JPY currency pair reveals that the PPO_T, DDQN_T, and PPO_NN strategies outperformed the perfect annual strategy benchmark in final balance as well as the buy-and-hold approach. These three models also surpassed benchmarks in risk-adjusted terms. However, the DDQN_NN strategy underperformed compared to these benchmarks. The PPO_T strategy stands out as the top performer, achieving a final balance of 18,120.64 and an impressive annualised return of 12.62%. It has a Sharpe ratio of 1.752, indicating excellent risk-adjusted returns This strategy particularly excelled in 2022 and 2023, capitalizing on favourable market conditions.

The DDQN_T strategy also demonstrated strong performance with a final balance of $16,105.72 and an annualised return of 10%. It achieved a Sharpe ratio of 1.252 and experienced a maximum drawdown of -8.37%. This strategy executed 257 trades with an average position duration of 5.45 days, the highest among all models, indicating a longer-term trading approach. The PPO_NN strategy achieved a final balance of $13,980.63 with an annualised return of 6.93% and a Sharpe ratio of 1.122, suggesting good risk-adjusted returns. It had a maximum drawdown of -5.76%, reflecting lower risk exposure compared to other strategies. This strategy engaged in 273 trades with an average position duration of 3.08 days and a win rate of 58.9%.

On the other hand, the DDQN_NN strategy exhibited moderate performance with a final balance of 11,960.99 and an annualised return of 3.64% with Sharpe ratio of 0.554, reflecting a significant but manageable level of risk. Overall, the PPO_T strategy emerged as the most effective, followed by DDQN_T and PPO_NN, each balancing profitability and risk differently. The DDQN_NN strategy, while balanced, did not reach the top-tier performance metrics of the PPO strategies.

### 4.3. USD-JPY

Next the presents are our results for the last FOREX currency pair - USD/JPY. According to Bank for International Settlements (2019), the USD/JPY currency pair accounted for approximately 13% of all daily traded volumes in the FOREX market. The USD/JPY exchange rate indicates how many Japanese Yen are needed to purchase one U.S. dollar. The USD/JPY pair is the second most traded currency pair in the global foreign exchange market. In Table 2, presented are performance results of our study while Figure 7 shows Profit and Losses i.e. total balance of strategy over time and Figure 8 represents maximum drawdown over time.

Table 3. Performance measures for DDQN-NN, DDQN-T, PPO-NN, PPO-T and two benchmarks over 2019-2023 years for the USD/JPY.

| USDJPY | Final balance | Provision Sum | Total trades | CAGR | Annualised std. | Sharpe ratio | Sortino Ratio | Maximum Drawdown | Maximum Drawdown Duration | Average Position Duration | Win Rate | In Long | In Short | Out of the market |
|---|---|---|---|---|---|---|---|---|---|---|---|---|---|---|
| **DDQN_NN** | 14 590.73 | -463.68 | 393 | 7.84% | 4.95% | 1.585 | 1.549 | -3.15% | 157 | 1.94 | 52.9% | 33.3% | 15.6% | 51% |
| **DDQN_T** | 16 092.60 | -377.97 | 300 | 9.98% | 6.61% | 1.511 | 1.601 | -7.66% | 184 | 3.12 | 56% | 39.4% | 20.7% | 39.8% |
| **PPO_NN** | 15 339.61 | -553.49 | 431 | 8.93% | 6.62% | 1.348 | 1.717 | -8.12% | 185 | 2.15 | 54% | 23.6% | 35.2% | 41.1% |
| **PPO_T** | 22 210.52 | -441.02 | 307 | 17.3% | 7.93% | 2.179 | 3.10 | -5.28% | 130 | 4.01 | 57.3% | 40.7% | 38.4% | 20.7% |
| | | | | | | **Benchmarks:** | | | | | | | | |
| **Buy and hold** | 12 887.18 | -1.0 | 1 | 6.11% | 9.27% | 0.742 | 0.915 | -15.1% | 777 | 1557 | 100% | 100% | 0% | 0% |
| **"perfect" annual strategy** | 14 485.73 | -2.05 | 2 | 9.05% | 9.18% | 0.943 | 1.189 | -15.1% | 315 | 778.5 | 100% | 60% | 40% | 0% |

*Source: Own study on the 2019-2023 daily data for USD/JPY (1557 observations) with starting capital of 10 000. DDQN is Double Deep Q-Network, PPO is Proximal Policy Optimisation, NN- fully connected Neural Netowork and T – transformer Network. The agents can take position with whole capital between long, short and out of the market. 'Perfect' annual strategy benchmark is a hypothetical trading strategy that predicts the optimal market position at the start of each year and holds that position, for the entire year based on the annual return forecast with perfect foresight.*

Figure 7. Balance changes for DDQN-NN, DDQN-T, PPO-NN, PPO-T and two benchmarks over 2019-2023 years for the USD/JPY.

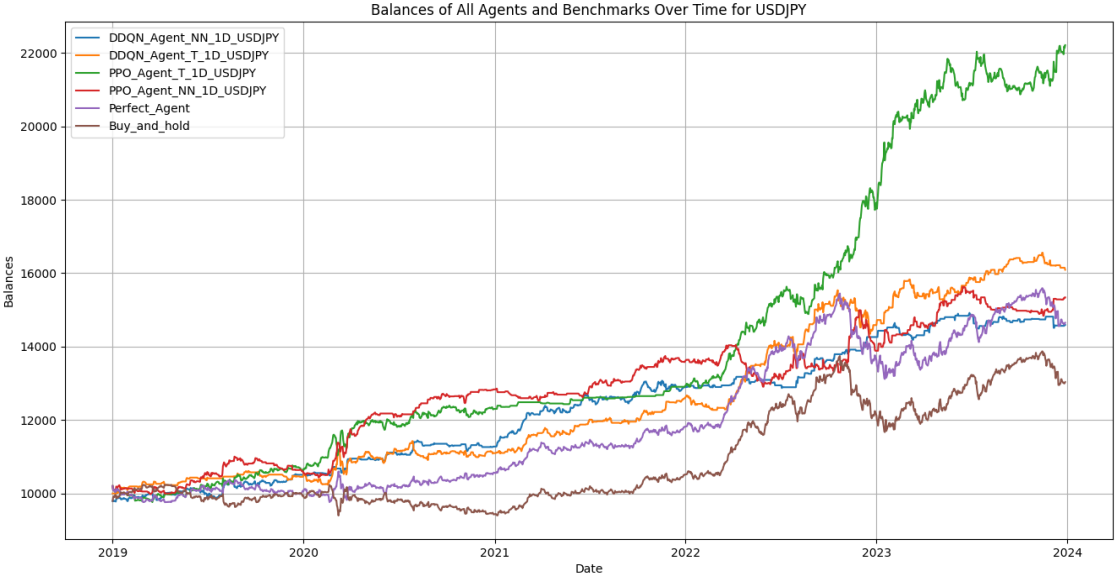

*Source: Own study on the 2019-2023 daily data for USD/JPY (1557 observations) with starting capital of 10 000. DDQN is Double Deep Q-Network, PPO is Proximal Policy Optimisation, NN- fully connected Neural Netowork and T – transformer Network. The agents can take position with whole capital between long, short and out of the market. 'Perfect' annual strategy benchmark is a hypothetical trading strategy that predicts the optimal market position at the start of each year and holds that position, for the entire year based on the annual return forecast with perfect foresight.*

Figure 8. Drawdowns for DDQN-NN, DDQN-T, PPO-NN, PPO-T and two benchmarks over 2019-2023 years for the USD/JPY.

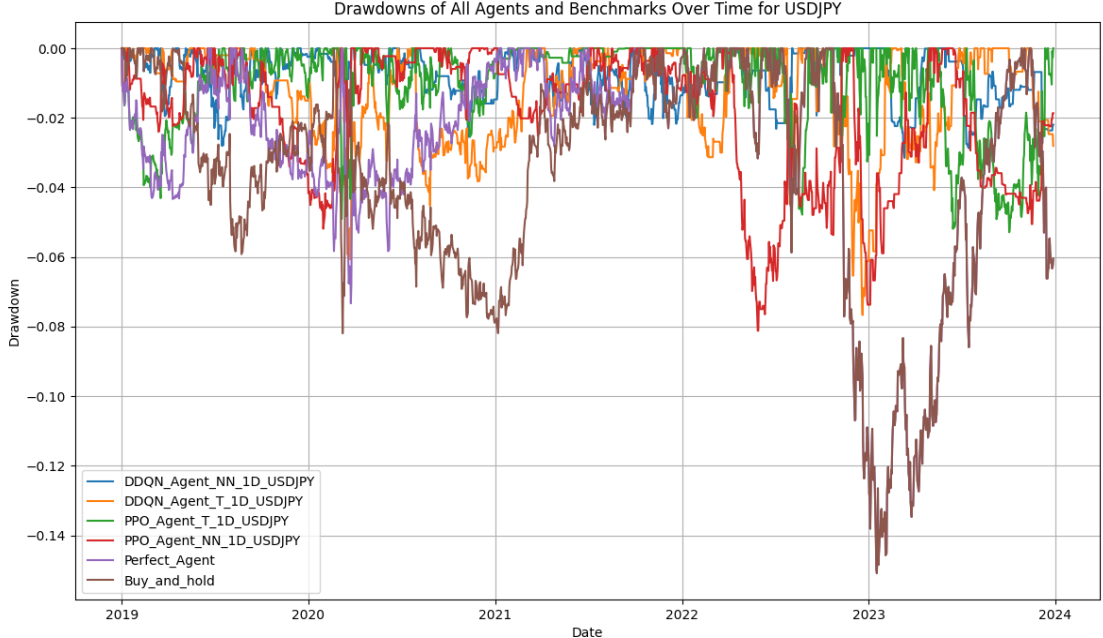

*Source: Own study on the 2019-2023 daily data for USD/JPY (1557 observations) with starting capital of 10 000. DDQN is Double Deep Q-Network, PPO is Proximal Policy Optimisation, NN- fully connected Neural Netowork and T – transformer Network. The agents can take position with whole capital between long, short and out of the market. 'Perfect' annual strategy benchmark is a hypothetical trading strategy that predicts the optimal market position at the start of each year and holds that position, for the entire year based on the annual return forecast with perfect foresight.*

All strategies demonstrated good results, but PPO_T stood out with an exceptional performance, achieving a final balance of 22,210.52 and a remarkable Sharpe ratio of 2.179, indicating significantly higher returns per unit of risk taken. PPO_T performed exceptionally well in 2022, capitalizing on favourable market conditions during that year. This aligns with the observed trend across all forex currencies, where PPO_T's strategy proved highly effective in 2022 and 2023. The strategy's frequent trading activity, with only 20.7% of the time out of the market, contributed to its high performance.

Other strategies, such as DDQN_NN, DDQN_T, and PPO_NN, also performed well, maintaining high values of time out of the market: 51%, 39.8%, and 41.1%, respectively. This conservative trading behaviour likely contributed to their ability to avoid adverse market movements and manage risk effectively. These strategies ended up with high values for risk-adjusted metrics. The drawdown analysis revealed that all strategies experienced relatively low drawdowns, with DDQN_T and PPO_NN having the largest drawdowns. However, all strategies had very short Maximum Drawdown Durations, indicating a quick recovery from losses. Additionally, PPO_T achieved the longest duration of holding positions, which further contributed to its exceptional performance. In summary, PPO_T emerged as the most effective strategy for trading USDJPY over the analysed period, offering the highest final balance and superior risk-adjusted returns. The analysis underscores the importance of balancing market exposure and risk, with PPO_T excelling due to its ability to navigate market conditions adeptly while maintaining high returns.

**4.4. S&P 500**

We continue the presentation of our results with the leading global stock market index, the S&P 500. The S&P 500 Index, introduced by Standard & Poor's in 1957, is widely regarded as the most significant benchmark of the U.S. equity market and is often considered a proxy for the overall health of the U.S. economy. According to S&P Dow Jones Indices, the S&P 500 accounts for approximately 80% of the U.S. stock market's total market capitalization, making it the most traded index in the world. The S&P 500 Index includes 500 of the largest publicly traded companies in the United States and reflects their market performance. Table 2 presents the performance results of our study, while Figure 5 shows the Profit and Losses, i.e., the total balance of the strategy over time, and Figure 6 represents the maximum drawdown over time.

Table 4. Performance measures for DDQN-NN, DDQN-T, PPO-NN, PPO-T and two benchmarks over 2019-2023 years for the S&P 500 index.

| S&P 500 | Final balance | Provision Sum | Total trades | CAGR | Annualised std. | Sharpe ratio | Sortino Ratio | Maximum Drawdown | Maximum Drawdown Duration | Average Position Duration | Win Rate | In Long | In Short | Out of the market |
|---|---|---|---|---|---|---|---|---|---|---|---|---|---|---|
| **DDQN_NN** | 20 009.45 | -1 657.70 | 454 | 22.43% | 22.55% | 0.993 | 1.225 | -15% | 157 | 2.35 | 51.3% | 51.7% | 33.5% | 14.7% |
| **DDQN_T** | 30 565.62 | -1 529.17 | 226 | 38.41% | 23.96% | 1.6 | 2.218 | -25.4% | 501 | 4.69 | 51.7% | 62.6% | 22% | 15.4% |
| **PPO_NN** | 11 567.96 | -697.29 | 274 | 4.3% | 23.67% | 0.183 | 0.193 | -41.6% | 972 | 3.35 | 54.7% | 59.8% | 13.4% | 26.7% |
| **PPO_T** | 40 010.83 | -1 526.78 | 251 | 49.76% | 23.05% | 2.158 | 2.651 | -26.1% | 181 | 4.53 | 58.6% | 73.9% | 16.9% | 9.1% |
| **Benchmarks:** | | | | | | | | | | | | | | |
| **Buy and hold** | 19 482.77 | -2.5 | 1 | 21.37% | 25.63% | 0.834 | 1.014 | -33.9% | 500 | 1253 | 100% | 100% | 0% | 0% |
| **"perfect" annual strategy** | 28 982.19 | -13.24 | 3 | 36.23% | 24.07% | 1.505 | 1.791 | -33.9% | 125 | 417.6 | 100% | 80% | 20% | 0% |

*Source: Own study on the 2019-2023 daily data for S&P 500 (1253 observations) with starting capital of 10 000. DDQN is Double Deep Q-Network, PPO is Proximal Policy Optimisation, NN- fully connected Neural Netowork and T – transformer Network. The agents can take position with whole capital between long, short and out of the market. 'Perfect' annual strategy benchmark is a hypothetical trading strategy that predicts the optimal market position at the start of each year and holds that position, for the entire year based on the annual return forecast with perfect foresight.*

Figure 9. Balance changes for DDQN-NN, DDQN-T, PPO-NN, PPO-T and two benchmarks over 2019-2023 years for the S&P 500.

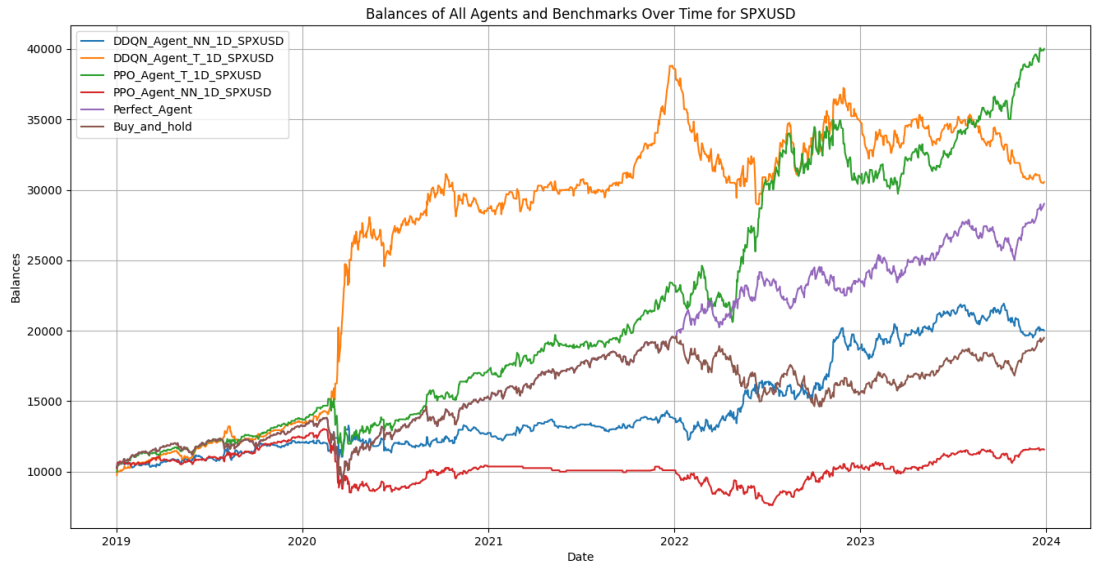

*Source: Own study on the 2019-2023 daily data for S&P 500 (1253 observations) with starting capital of 10 000. DDQN is Double Deep Q-Network, PPO is Proximal Policy Optimisation, NN- fully connected Neural Netowork and T – transformer Network. The agents can take position with whole capital between long, short and out of the market. 'Perfect' annual strategy benchmark is a hypothetical trading strategy that predicts the optimal market position at the start of each year and holds that position, for the entire year based on the annual return forecast with perfect foresight.*

Figure 10. Drawdowns for DDQN-NN, DDQN-T, PPO-NN, PPO-T and two benchmarks over 2019-2023 years for the S&P 500.

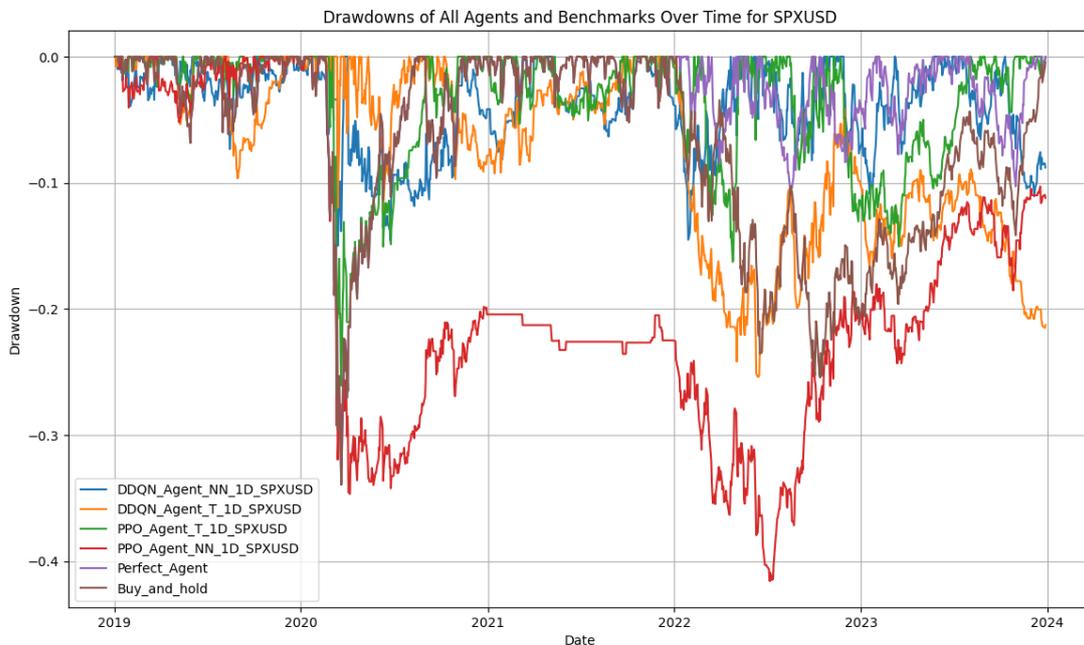

*Source: Own study on the 2019-2023 daily data for S&P 500 (1253 observations) with starting capital of 10 000. DDQN is Double Deep Q-Network, PPO is Proximal Policy Optimisation, NN- fully connected Neural Netowork and T – transformer Network. The agents can take position with whole capital between long, short and out of the market. 'Perfect' annual strategy benchmark is a hypothetical trading strategy that predicts the optimal market position at the start of each year and holds that position, for the entire year based on the annual return forecast with perfect foresight.*

The analysis of various trading strategies on the S&P 500 over five years, as depicted in the provided Table 4 and Figures 9 and 10, offers critical insights into their performance and key findings. Two of our agents (PPO_T and DDQN_T) managed to outperform the 'perfect' annual strategy in terms of final capital, while DDQN_NN barely surpassed the Buy and Hold benchmark. A key finding is that DDQN_T was the only agent to profit enormously during the COVID-19 crash (early 2020). This initial success was followed by three years of average returns, suggesting that while the strategy can capitalize on volatile events, its long-term consistency may be less robust. Conversely, PPO_T exhibited a steady increase over the examined period, suggesting excellent adaptability and resilience of the strategy to varying market conditions. Additionally, PPO_T's lower Maximum Drawdown Duration of 181 days, compared to DDQN_T's 501 days, indicates better risk management and quicker recovery from losses. This highlights PPO_T's ability to maintain stability and consistent growth even during turbulent times.

DDQN_NN and PPO_NN performed poorly overall, with PPO_NN significantly underperforming throughout the five-year period. In contrast, DDQN_NN, although slow, demonstrated steady gains, ultimately beating the Buy and Hold benchmark in the category of risk-adjusted returns, as indicated by higher Sharpe and Sortino ratios. Another important observation is that all agents had a relatively low value of staying out of the market. The highest out-of-market percentage was 26.7% by PPO_NN, which also had the poorest performance. The evaluation of trading strategies on the S&P 500 reveals that PPO_T stands out as the most effective approach, offering the highest final balance and annualized returns with a strong Sharpe ratio, despite experiencing moderate drawdowns.

**4.5. Bitcoin**

Last but not least we will present results for Bitcoin, the most known and traded cryptocurrency in the digital currency market. According to market analysis, Bitcoin consistently dominates trading volumes and market capitalization, representing approximately 50% of the total cryptocurrency market capitalization. This makes Bitcoin the most liquid and widely traded digital asset globally (Coin Gecko, 2024). The Bitcoin price indicates how much one unit of Bitcoin is worth in terms of fiat currency, in our case the U.S. dollar. Crucial to acknowledge is that Bitcoin serves as a benchmark for the overall health and performance of the cryptocurrency market.

Table 5. Performance measures for DDQN-NN, DDQN-T, PPO-NN, PPO-T and two benchmarks over 2019-2023 years for the Bitcoin cryptocurrency notated in USD.

| BTC | Final balance | Provision Sum | Total trades | CAGR | Annualised std. | Sharpe ratio | Sortino Ratio | Maximum Drawdown | Maximum Drawdown Duration | Average Position Duration | Win Rate | In Long | In Short | Out of the market |
|---|---|---|---|---|---|---|---|---|---|---|---|---|---|---|
| **DDQN_NN** | 38 822.79 | -15 373.38 | 335 | 31.16% | 61.8% | 0.504 | 0.704 | -78.7% | 601 | 4.93 | 46.2% | 65% | 25.6% | 9.4% |
| **DDQN_T** | 50 038.35 | -2 518.51 | 64 | 37.99% | 37.8% | 1.01 | 0.86 | -59.8% | 1087 | 8.23 | 45.3% | 28.8% | 0% | 71.1% |
| **PPO_NN** | 9 410.26 | -5 120.4 | 263 | -1.2% | 53.2% | -0.022 | -0.025 | -85.3% | 1068 | 5.04 | 52% | 18.8% | 53.8% | 27.3% |
| **PPO_T** | 140 551.23 | -25 460.93 | 388 | 69.65% | 63.7% | 1.093 | 1.593 | -55.1% | 714 | 4.14 | 51.5% | 58% | 30.1% | 11.8% |
| | | | | | **Benchmarks:** | | | | | | | | | |
| **Buy and hold** | 106 739.80 | -10.0 | 1 | 60.6% | 68% | 0.89 | 1.191 | -76.6% | 783 | 1824 | 100% | 100% | 0% | 0% |
| **"perfect" annual strategy** | 499 023.89 | -32.61 | 3 | 118.6% | 63.4% | 1.871 | 2.47 | -61.9% | 483 | 608 | 100% | 80% | 20% | 0% |

*Source: Own study on the 2019-2023 daily data for BTC (1824 observations) with starting capital of 10 000. DDQN is Double Deep Q-Network, PPO is Proximal Policy Optimisation, NN- fully connected Neural Netowork and T – transformer Network. The agents can take position with whole capital between long, short and out of the market. 'Perfect' annual strategy benchmark is a hypothetical trading strategy that predicts the optimal market position at the start of each year and holds that position, for the entire year based on the annual return forecast with perfect foresight.*

Figure 11. Balance changes, on the logarithmic for DDQN-NN, DDQN-T, PPO-NN, PPO-T and two benchmarks over 2019-2023 years for the Bitcoin currency notated in USD.

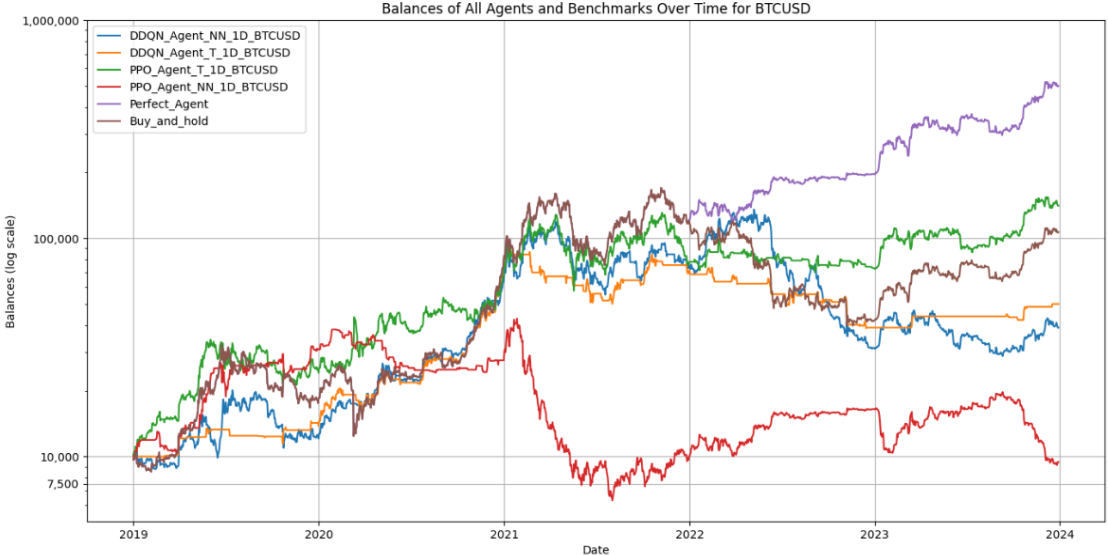

*Source: Own study on the 2019-2023 daily data for BTC (1824 observations) with starting capital of 10 000. DDQN is Double Deep Q-Network, PPO is Proximal Policy Optimisation, NN- fully connected Neural Netowork and T – transformer Network. The agents can take position with whole capital between long, short and out of the market. 'Perfect' annual strategy benchmark is a hypothetical trading strategy that predicts the optimal market position at the start of each year and holds that position, for the entire year based on the annual return forecast with perfect foresight.*

Figure 12. Drawdowns for DDQN-NN, DDQN-T, PPO-NN, PPO-T and two benchmarks over 2019-2023 years for the Bitcoin currency notated in USD.

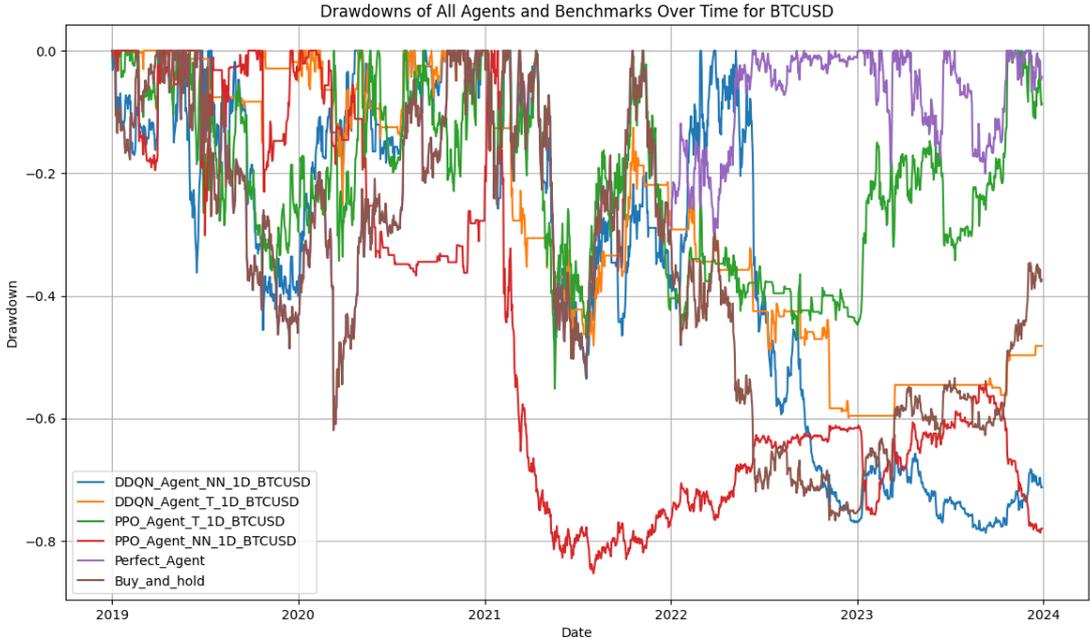

*Source: Own study on the 2019-2023 daily data for BTC (1824 observations) with starting capital of 10 000. DDQN is Double Deep Q-Network, PPO is Proximal Policy Optimisation, NN- fully connected Neural Netowork and T – transformer Network. The agents can take position with whole capital between long, short and out of the market. 'Perfect' annual strategy benchmark is a hypothetical trading strategy that predicts the optimal market position at the start of each year and holds that position, for the entire year based on the annual return forecast with perfect foresight.*

The analysis of various trading strategies on Bitcoin (BTC/USD) over five years, as depicted in Table 5 and Figures 11 and 12, offers critical insights into their performance and key findings. None of the agents beat the 'perfect' annual strategy. However, only PPO_T managed to outperform the Buy and Hold strategy in terms of final balance. Both DDQN_T and PPO_T surpassed the Buy and Hold benchmark in risk-adjusted terms, although they did not overcome the perfect annual strategy.

The DDQN_NN strategy showed substantial performance with a final balance of $38,822.79 and an annualized return of 31.16%. The Sharpe ratio of 0.504 indicates moderate risk-adjusted returns. This strategy experienced a high maximum drawdown of -78.7%, reflecting significant risk exposure. Behaviorally, DDQN_NN executed 335 trades with an average position duration of 4.93 days. Suprisly, the win rate was below of 50% at 46.2%. However, DDQN_NN manages to be highly net profitable. DDQN_NN highly underperformed in 2022 and 2023. The DDQN_T strategy demonstrated impressive performance with a final balance of $50,038.35 and an annualized return of 37.99%. The Sharpe ratio of 1.01 indicates favorable risk-adjusted returns and in the risk-adjusted terms DDQN_T outperformed Buy and Hold benchmark. Behaviorally, DDQN_T executed only 64 trades with an average position duration of 8.23 days. It has also negative win rate 45.3%. The strategy held positions long 28.8% of the time and was out of the market for 71.1% of the time and never decided to short the market.

The PPO_NN strategy underperformed relative to other strategies, achieving a final balance of $9,410.26 with an annualized return of -1.2% and the strategy experienced a maximum drawdown of -85.3%. In 2021 PPO_NN has very poor performance while all other strategies manges to substantially achive gains. The PPO_NN strategy's poor performance metrics highlight its inefficacy in managing risks and generating consistent profits. The PPO_T strategy outperformed all other strategies with a final balance of $140,551.23 and an impressive annualized return of 69.65%. The Sharpe ratio of 1.093 indicates good risk-adjusted returns. The strategy experienced a maximum drawdown of -55.1%, which, while high, is acceptable given high volatile nature of Bitcoin. Even if most of gain were achived in first 2 years of the examined period.

## 4.6. Results summary

The analysis of various trading strategies across multiple assets, including FOREX currency pairs, S&P 500 and Bitcoin, reveals a clear hierarchy in performance. The PPO_T strategy consistently outperformed other strategies, followed by DDQN_T, DDQN_NN and lastly, PPO_NN. PPO_T demonstrated superior performance across all assets. For EUR/USD, which was the only asset it had second best performance. Worth to point out is fact that for Bitcoin none of agents beat the 'perfect' annual strategy, while only PPO_T beaten buy and hold benchmark.

In the FOREX market, DRL agents performed exceptionally well, especially transformer networks, which achieved enormous gains in 2022 and 2023. Perfect gains were achieved by capitalizing on short 5-10 day trends rather than a high number of swings or longer trends without corrections. For the S&P 500, the DDQN_T strategy was the only one with strong performance during the COVID-19 period, achieving massive gains. PPO_T, on the other hand, mitigated some losses with an out-of-market approach on certain days, resulting in lower losses compared to the overall buy-and-hold strategy during that period. All agents performed well in both bear and bull markets, indicating that DRL can perform effectively regardless of market conditions. Despite these successes, the poor performance of DRL agents in Bitcoin trading can be attributed to several factors. A critical issue was the shorter BTC training period, which likely contributed to suboptimal results. DRL requires extensive data to train effectively and insufficient data limits the agent's ability to learn and adapt to market dynamics.

Overall, the PPO_T strategy emerged as the most effective, followed by DDQN_T and DDQN_NN, each balancing profitability and risk differently. The PPO_NN strategy, while balanced in some cases, did not reach the top-tier performance metrics of the other strategies.

The gamma parameter in Temporal Difference (TD), as mentioned before, plays a crucial role in balancing between immediate and future rewards. Higher gamma values suggest greater focus on long term profits while low value would result in higher focus on short term gains. In our case this would mean that high value of gamma should result in longer durations for holding positions. Conversely, a gamma parameter of 0 focuses the agent solely on immediate returns, looking only at the reward at next time step ($t+1$). In such a scenario, the agent is likely to engage in very frequent trading, as it continuously seeks to capitalise on immediate opportunities without regard for the future. As for our results we can see that the value of 0.75 has not indicated long term trading we were aiming for.

Additionally, we have tested higher values of gamma of 0.9 and 0.95. Suppressively, agents had lower trading duration times as well as performance. This could be to attributed to two primary factors. Firstly, an excessively high gamma parameter may distort reward estimates. When gamma is set too high, the rewards from distant future events are discounted minimally, causing the agent to perceive future rewards as nearly equivalent. This lack of differentiation can disturb the agent's learning process, as it becomes challenging to distinguish between different rewards. For example, with a gamma value of 0.95, the reward after 13 time steps is discounted by only half, resulting in a blurred perception of future rewards. Secondly, the issue may arise from the input state configuration. With an input comprising 20 previous observations, the agent might struggle to make accurate predictions far into the future. This suggests that the gamma parameter is somehow linked to the length of the historical data used by the agent.

Moreover, the strategic non-engagement capability of DRL can be particularly beneficial during market anomalies or black swan events, where traditional models often fail in predicting future outcomes. During such unprecedented times, characterized by conditions never before encountered by the DRL agent before, the agent is likely to opt out of the market. This decision stems from its learned behaviour of exercising caution in uncertain never faced before conditions. By leveraging its trial-and-error learning process, the DRL agent identifies periods of high uncertainty and decides when to stay inactive, effectively preventing significant losses.

To conclude, the results provide a robust answer to the main research question of this study: Deep Reinforcement Learning (DRL), as an advanced form of artificial intelligence, can effectively conduct financial trading by autonomously identifying and exploiting patterns and relationships within complex, high-dimensional data. The PPO model on a transformer network exemplifies this capability, with Sharpe ratios across assets of 0.847, 1.752, 2.179, 2.158, and 1.093, demonstrating its effectiveness. However, it is crucial to highlight that successful DRL application in trading requires not only advanced algorithms like PPO and robust neural network architectures such as transformers but also an extensive and high-quality dataset for training. The poor performance observed in Bitcoin trading underscores the importance of a sufficiently long training period and a well-calibrated gamma parameter. These elements are essential to ensure that the DRL agent can effectively balance immediate and future rewards, adapt to market dynamics, and achieve optimal trading performance.

# CHAPTER V
## Discussion and future work

### 5.1. Methodological Challenges in Applying DRL to Financial Markets

The results of this paper look very promising. This raises an intriguing question: Could DRL be the new 'holy grail' of trading? To address this, it is crucial to analyse specific methodological issues associated with reinforcement learning in the context of trading financial markets.

First is the Temporal Difference. In TD learning and its variations, future rewards are discounted back to the present action, which can be problematic in the context of trading. For example, consider a sequence of actions: *long*, *long*, *short*, *short*. The TD algorithm assigns a discounted value of estimated future rewards from the *short* positions back to the initial *long* actions. The reward from the future *short* is discounted and partially attributed to the initial *long* position as we discount 'policy' choices. However, this approach can be problematic because financial markets are highly dynamic and influenced by numerous unpredictable factors. The discounted reward from a future action may not accurately reflect the true value of the current decision, leading to suboptimal strategies.

Secondly, a critical aspect to consider is the assumption of reinforcement learning models is that the agent's decisions significantly influence the environment. The RL model assumes that its decision changes the future state $s_{t+1}$ that he would end up after the action *a*, which is inherently not true in our case. In the realm of algorithmic trading, this influence is generally limited only to the positions the agent holds in the market and is freely changing. Crucially, individual transactions, typically of minimal volume, do not impact broader market prices substantially. And even in hypothetical scenarios where an agent could execute transactions with billions in volume, there remains a significant problem: we are training models on historical prices. Such historical data cannot feasibly replicate the impact an agent's decision would have on the market. To sum up, the agent thinks that his actions are changing prices which is definitely not true.

The third issue concerns the methodology of policy-based models, such as Proximal Policy Optimisation (PPO). These models produce stochastic probabilities for each action, introducing an element of randomness in decision-making. In out-of-sample testing, the agent selects actions with the highest probability to maximise expected profits. However, this approach is inherently flawed in the real-world, as the agent is sometimes choosing actions with 90% probability, and sometimes with only 35%. It is important to note that this issue is not present in value-based methods like Double Deep Q-Network (DDQN). In these frameworks, the agent naturally chooses the action associated with the highest Q-value, which directly corresponds to the best action.

Last but not least, we must address the issue with eligibility traces. This concept, pivotal in complex, long-duration games, involves tracing rewards backward to earlier actions that significantly influence outcomes much later in the game. For example, an action such as opening shortcut doors early in a video game can yield considerable advantages as the game progresses. The reward for this action is then traced back and a discounted premium is awarded to the decision, acknowledging its long-term impact. In the context of the stock market, however, the utility of eligibility traces is less straightforward. Traders can adjust their positions frequently at each step, which contrasts sharply with the crucial decisions made early in long games. While there are costs associated with maintaining positions, such as provisions for keeping a position open overnight, these are generally negligible compared to the potential gains from significant price movements on highly liquid markets. Thus, while eligibility traces offer a sophisticated method for attributing value to actions based on their eventual outcomes in certain scenarios, their effectiveness in the fast-paced, highly fluid environment of the stock market is debatable.

## 5.2. Risk control

The primary advantage of Deep Reinforcement Learning (DRL) in the context of stock market trading lies in its capacity for managing risk of trading strategy. One of the features of DRL is to strategically choose not to engage in the market. This characteristic is pivotal for achieving superior performance. Unlike most of traditional methods, where decision-making is binary (either buy or sell), DRL agents exhibit a preference for remaining inactive about approximately 30% of the time, thereby avoiding potentially unprofitable trades.

In contrast, supervised learning lacks a direct mechanism for staying out of the market. Decisions are based on predefined rules, and programmers must introduce additional trading guidelines for scenarios where the probabilities of long and short positions are nearly equal, typically around 50%. This process can be less effective and rely on additional rules defined by human interventions and optimised thresholds, which may not adapt well to dynamic market conditions.

DRL, on the other hand, learns to avoid trading during unfavourable conditions. Through continuous interaction with the market environment, the agent is able to recognize patterns and situations that are likely to result in losses. It then opts to stay out of the market during these times. This adaptive learning process allows the agent to capitalise on favourable market situations while minimising exposure to risk. This behaviour demonstrates a significant edge over classical supervised learning approaches.

This adaptive behaviour of DRL agents can be integrated into the framework of the Efficient Market Hypothesis (EMH). Traditionally, the EMH suggests that it is impossible to 'beat the market' consistently because stock prices always incorporate and reflect all relevant information. However, DRL introduces a nuanced perspective: perhaps the way to achieve superior performance is not through perfect price prediction but through strategic non-engagement. By knowing when to avoid trading, DRL can potentially exploit inefficiencies that arise from market volatility and investor behaviour.

An illustrative example of the power of strategic non-engagement is seen in the historical performance of the S&P 500. If an investor had managed to avoid the 10 worst trading days since the year 2003, the overall return of a buy-and-hold strategy would have increased by 2.5 times. Which, it is worth to note, is an outstanding result. This highlights the importance of knowing when to trade and, more importantly, when to stay out of the market. The ability to avoid these significant losses underscores the critical role of risk management in achieving long-term financial success.

Moreover, the strategic non-engagement capability of DRL can be particularly beneficial during market anomalies or black swan events, where traditional models often fail in predicting future outcomes. During such unprecedented times, characterized by conditions never before encountered by the DRL agent, the agent is likely to opt out of the market. This decision stems from its learned behaviour of exercising caution in uncertain never faced before conditions. By leveraging its trial-and-error learning process, the DRL agent identifies periods of high uncertainty and decides when to stay inactive, effectively preventing significant losses.

## 5.3. Future work

To further advance the application of Deep Reinforcement Learning (DRL) in trading, several promising avenues for future research can be explored. This includes enhancing risk control mechanisms and addressing the stochastic nature of policy-based methods through innovative approaches, such as adjustments in the size of positions, the use of Hierarchical

Reinforcement Learning (HRL) and the incorporation of hyperparameter optimisation techniques.

Moving forward, optimisation approaches can be enhanced by the incorporation of techniques such as random, grid or Bayesian search. Those methods would explore a range of hyperparameter values in order to identify the optimal configuration for the DRL models. Those could significantly improve the robustness and effectiveness of agents in trading environments.

The next potential upgrade is to empower the agent to adjust its position dynamically. Instead of merely deciding to buy, sell, or stay out of the market, the agent would independently determine the amount of capital to invest (or the percentage of capital) based on the current state of the environment. This capability allows the agent to manage risk with even greater precision, akin to the decision to stay out of the market. By permitting the agent to select from a continuum of investment sizes within a specified range (e.g., between -1 and 1), the model achieves finer control over its risk exposure. This approach is particularly well-suited to the stochastic nature of policy-based algorithms. The agent can offer a distribution of probabilities across the investment range, thereby enabling more accurate risk management. However, implementing this method with value-based models would be infeasible, as they depend on selecting discrete actions based on their associated Q-values.

Another promising implementation that would take DRL to the next level is Hierarchical Reinforcement Learning (HRL). In short, this branch of reinforcement learning enhances DRL models by introducing multi-agent frameworks. In this approach, the master agent manages the allocation of capital among subordinate agents, each specialising in trading specific assets or in market conditions. The master agent operates at the top of the hierarchy, overseeing decisions and resource allocation based on performance of subagents. However, the implementation of HRL poses unique challenges, including designing effective communication protocols among agents, managing conflicts of interest, and optimising the hierarchical structure to maximise overall performance. This modification would add further complexity to the already intricate model of trading.

# CONCLUSIONS

In this study, we investigated the efficacy of Artificial Intelligence, specifically Deep Reinforcement Learning (DRL), in trading financial markets. We focused on the potential of two advanced DRL algorithms: Double Deep Q-Network (DDQN) and Proximal Policy Optimisation (PPO). Our research encompassed a range of financial assets, including three currency pairs (EUR/USD, EUR/JPY, USD/JPY), the S&P 500 index and Bitcoin. The algorithms were tested on daily data using a moving forward optimisation method with a one-year window, covering the period from 2019 to 2023.

Our methodology involved the application of DRL algorithms to autonomously identify and exploit patterns within high-dimensional market data. The algorithms used were DDQN, which mitigates the overestimation bias common in Q-learning by utilizing two separate networks for action selection and evaluation, and PPO, a policy-based method that employs an actor-critic framework with a clipping mechanism to maintain stability and improve learning efficiency. Deep Learning was used to approximate Reinforcement Learning (RL) algorithms. We tested these algorithms with two types of neural network architectures: Fully Connected Neural Networks and Transformer Networks, to evaluate their performance across different market conditions.

Our research findings robustly support the research hypothesis that Deep Reinforcement Learning (DRL) can effectively conduct financial market trading by autonomously identifying and exploiting patterns within complex data sets. The empirical results indicate that the PPO algorithm combined with Transformer Networks outperformed all other methods (average Sharpe ratio between our studies of 1.61), demonstrating superior effectiveness in trading financial markets. This was followed by DDQN with Transformer Networks, which also showed strong performance. DDQN with Fully Connected Neural Networks came next, providing moderate results. Surprisingly, the worst performance was observed with PPO on Fully Connected Neural Networks, which generally underperformed in most scenarios. These findings highlight the significant advantages of Transformer Networks in DRL applications. The attention mechanism in Transformer Networks, with its ability to capture intricate and sequential patterns within financial data, proved highly beneficial. This sequential approach to data processing allowed the Transformer Networks to better understand and respond to market dynamics compared to Fully Connected Neural Networks.

While PPO showed overall better performance than DDQN, the poor results of PPO on Fully Connected Neural Networks suggest that the choice of neural network architecture plays a crucial role in the success of DRL algorithms. The remarkable performance of PPO with Transformer Networks emphasizes the importance of leveraging advanced architectures to enhance Reinforcement Learning. Conversely, the weak performance of PPO with Fully Connected Neural Networks shows that even strong RL algorithms can struggle if not matched with the right network architecture.

One of the significant advantages of DRL in trading is its ability to manage risk effectively by choosing not to engage in the market during unfavourable conditions. This characteristic led to improved performance, enhancing the overall Sharpe ratio compared to traditional buy-and-hold strategies. Avoiding market engagement during adverse conditions might be a key strategy to outperform benchmarks, potentially challenging the efficient-market hypothesis by exploiting market inefficiencies.

DRL represents a powerful tool for algorithmic trading, capable of identifying and exploiting market conditions through trial-and-error learning. The ongoing advancements in AI, Machine Learning and Deep Reinforcement Learning will likely continue to enhance the effectiveness and reliability of these trading systems, paving the way for more sophisticated and adaptive financial technologies. As research and development in this field progress, we can expect DRL to play a crucial role, offering innovative solutions for managing risk and maximizing returns. Additionally, future studies could focus on developing DRL agents that autonomously decide the amount to invest, further refining the decision-making process and enhancing risk-awareness trading strategies. This approach could provide a significant step forward in creating even more effective and adaptive financial trading systems.

# LIST OF APPENDICES

**List of shorts**

| | | |
|---|---|---|
| CAGR | - | Compound annual growth rate |
| DDQN | - | Double Deep Q-Network |
| DL | - | Deep Learning |
| DRL | - | Deep Reinforcement Learning |
| DRL | - | Deep Reinforcement Learning |
| DQN | - | Deep Q-Network |
| EMH | - | Efficient Market Hypothesis |
| FOREX | - | Foreign Exchange Market |
| HRL | - | Hierarchical Reinforcement Learning |
| NN | - | Neural Network |
| PnL | - | Profits and Losses |
| PPO | - | Proximal Policy Optimisation |
| SARSA | - | State–Action–Reward–State–Action |
| T | - | Transformer Network |
| TD | - | Temporal Difference |
| $A$ | - | Advantage |
| $A^{GAE}$ | - | Generalised Advantage Estimation |
| $a$ | - | Action |
| $\pi$ | - | Policy |
| $r$ | - | Reward |
| $s$ | - | State |

**List of Tables**

**List of Figures**